\RequirePackage{fix-cm}
\RequirePackage{rotating}
\documentclass[smallextended]{svjour3}       
\smartqed  

\usepackage{graphicx}

\usepackage[nocompress]{cite}
\usepackage{makecell}
\usepackage{amsmath}
\usepackage{algorithmic}
\usepackage{array}
\usepackage[round]{natbib}
\usepackage{subcaption}
\usepackage{url}
\usepackage{listings}
\usepackage{booktabs}
\usepackage{array}
\newcolumntype{L}[1]{>{\raggedright\let\newline\\\arraybackslash\hspace{0pt}}m{#1}}
\newcolumntype{C}[1]{>{\centering\let\newline\\\arraybackslash\hspace{0pt}}m{#1}}
\newcolumntype{R}[1]{>{\raggedleft\let\newline\\\arraybackslash\hspace{0pt}}m{#1}}
\usepackage{multirow}
\usepackage{color,soul}
\usepackage[utf8]{inputenc}
\usepackage[T1]{fontenc}
\usepackage{xcolor}
\usepackage{tabularx}
\usepackage{framed}
\usepackage[normalem]{ulem}

\newcommand{\del}[1]{}
\newcommand{\add}[1]{{{#1}}}

\newcommand{\cfr}{\emph{cfr.}~}
\newcommand{\ie}{\emph{i.e.,}~}
\newcommand{\eg}{\emph{e.g.,}~}

\usepackage{pdfcomment}
\usepackage{marginnote}

\usepackage{multirow}

\makeatletter
\newcommand{\revDX}[1]{%
\marginnote{ \textbf{\textcolor{blue}{Revision #1}}}
}
\newcommand{\revSX}[1]{%
\begingroup%
\reversemarginpar\marginnote{\textbf{\textcolor{blue}{Revision #1}}}%
\endgroup
}
\newcommand{\rev}[1]{%
\if@firstcolumn
\revSX{#1}%
\else
\revDX{#1}%
\fi
}
\makeatother

\begin{document}

\title{A Comprehensive Study on Software Aging\\across Android Versions and Vendors}


\author{Domenico Cotroneo         \and
        Antonio Ken Iannillo	  \and
        Roberto Natella \and
        Roberto Pietrantuono
}


\institute{Domenico Cotroneo, Antonio Ken Iannillo, Roberto Natella, Roberto Pietrantuono \at
              Università degli Studi di Napoli Federico II\\
              via Claudio 21, 80125, Napoli, Italy \\
              \email{\{cotroneo, antonioken.iannillo, roberto.natella, roberto.pietrantuono\}@unina.it}
}

\date{Received: date / Accepted: date}

\maketitle

\begin{abstract}
This paper analyzes the phenomenon of software aging -- namely, the gradual performance degradation and resource exhaustion in the long run -- in the Android OS. The study intends to highlight if, and to what extent, devices from different vendors, under various usage conditions and configurations, are affected by software aging and which parts of the system are the main contributors. The results  demonstrate that software aging systematically determines a gradual loss of responsiveness perceived by the user, and an unjustified depletion of physical memory. The analysis reveals differences in the aging trends due to the workload factors and to the type of running applications, as well as differences due to vendors' customization. 
Moreover, we analyze several system-level metrics to trace back the software aging effects to their main causes. We show that bloated Java containers are a significant contributor to software aging, and that it is feasible to mitigate aging through a micro-rejuvenation solution at the container level.

\end{abstract}

\section{Introduction}
\label{sec:intro}

With mobile devices becoming crucial for our everyday activities, the need for designing reliable and high-performance software for smartphones is well recognized. At the same time, the numerous new functions required to satisfy the emerging customers' needs, along with the short time-to-market, greatly impact the size, complexity and, ultimately, the quality of the delivered software. This turns into frequent software-related failures, ranging from degraded performance to the device hang or even crash. 

A common problem, whose impact on end-user quality perception is often underestimated by engineers, is \textbf{software aging} \citep{huang1995software}. 
Software affected by the so-called \textit{aging-related bugs} (ARBs) suffers from the gradual accumulation of errors that induces to progressive performance degradation, and eventually to failure \citep{grottke2007fightingbugs,machida2012aging}. Due to such a ``subtle'' depletion, ARBs are difficult to diagnose during testing; they appear only after a long execution and under non-easily reproducible triggering and propagation conditions. 
Typical examples include memory leakages, fragmentation, unreleased locks, stale threads, data corruption, and numerical error accumulation, which gradually affect the state of the environment (e.g., by consuming physical memory unjustifiably). The typical solution is to figure out the temporal trend of the degradation, in order to act by preventive maintenance actions known as \textit{rejuvenation}, i.e., solutions to clean and restore the degraded state of the environment \citep{huang1995software,grottke2016recovery}. The problem is known to affect many software systems, ranging from business-critical to even safety-critical systems \citep{garg1998methodologyaging,silva2006soap,grottke2006agingwebserver,carrozza2010memory,grottke2010empirical,araujo2014software,cotroneo2014survey}. 

In this work, we focus on the \emph{Android OS}. Software aging in the Android OS can potentially affect the user experience of millions of mobile products: this OS currently dominates the smartphone market, and has been approaching 50 millions of physical lines of code\footnote{Computed with David A. Wheeler’s SLOCCount \citep{sloccount} on the entire Android Open Source Project (AOSP) Nougat 24 \citep{aosp}.}. We conducted an extensive experimental study to highlight potential aging phenomena, to understand the conditions when they occur more severely, and to diagnose their potential source. Indeed, we designed and performed a controlled experiment, grounding on a series of long-running tests, where devices from four different vendors (Samsung, Huawei, LG, and HTC) were stressed and monitored under various configurations.

Results revealed i) systematic trends of performance degradation, which manifest under all the tested Android versions, vendors, and configurations; ii) that software aging is exacerbated by the type of workload (as the degradation trends are more severe under applications from the Chinese market compared to the European one), and by vendor customizations; and  iii) that the performance degradation trends are not improving across different Android Versions (from Android 5 to 7): this emphasizes the need to invest more on software aging problems.

By correlating the aging trends to resource utilization metrics, we found that memory consumption is the main issue. In all the cases, the main aging issues can be confined to key processes of the Android OS that exhibit a systematic inflated memory consumption. Specifically, a correlation analysis reveals that the \emph{System Server} process plays a key role in determining the bad performance in terms of responsiveness, and thus should be the main target of a rejuvenation action. Such a result is corroborated by the further analysis conducted on Garbage Collection metrics, where we observed a significant and systematic increase of collection times of objects of the \emph{System Server} process. 
Moreover, we present a further experiment to provide more insights about the root causes of the aging, by applying a simple micro-rejuvenation mechanism on selected Java containers. The experiment shows that bloated Java containers indeed are a significant contributor to software aging, and that it is feasible to mitigate aging through a micro-rejuvenation solution at the container level.

\add{
This work is an extension of a previous initial analysis of software aging in the Android OS \citep{cotroneo2016software}. Previous studies in this field, including ours \citep{cotroneo2016software} and work from other researchers \citep{qiao2016empirical,weng2016analysis}, were limited to the analysis of an individual Android vendor and version of the OS; for example, our previous work has been focused on software aging in the Android OS version 5 (Lollipop) from the Huawei vendor. Android vendors introduce significant customizations (\emph{e.g.,} to improve the user experience and stand out against competition), it becomes important to consider many of them, in order to quantify the impact of customizations and the extent of the problem. Moreover, the Android OS has been evolving over several years, with many major versions that revised and extended the architecture of this OS. Several empirical studies showed that these differences among versions and among vendor customizations have a significant impact on both feature-richness and reliability, such as, in terms of compatibility issues \citep{wei2016taming,li2018cid,wei2018understanding,scalabrino2019data} and security vulnerabilities \citep{wu2013impact,gallo2015security,iannillo2017chizpurfle}.
}

\add{
For these reasons, it becomes important to consider how software aging impacts on different flavors of the Android OS. Therefore, this paper significantly extends the previous analysis by covering Android devices from four major \emph{vendors} (Samsung, Huawei, HTC, LG), by investigating the differences across them, and extends the analysis to three major \emph{versions} of the Android OS. Moreover, we here provide a more detailed analysis of the root causes of software aging, by tracing back the issue to bloated Java containers. 
}

In Section~\ref{sec:related}, we first survey the related literature on software aging and mobile device reliability. Section~\ref{sec:research_questions} provides an overview of the research questions addressed in this paper. Section~\ref{sec:methodology} describes our experimental methodology. Section~\ref{sec:results} presents the experimental results, which are discussed in the conclusive Section~\ref{sec:conclusion}.

\section{Related work}
\label{sec:related}

Software aging has been extensively studied since the early '90s \citep{huang1995software,cotroneo2014survey}. One important branch of research has been on \emph{analytical modeling} of systems under software aging, in order to find the optimal time at which to schedule software rejuvenation  \citep{pfening1996optimal,dohi2000statistical,vaidyanathan2005comprehensive,grottke2016recovery}, and to perform actions to extend the software lifetime, such as throttling the workload and allocating more resources \citep{machida2016lifetime}. Another branch of research has been on the \emph{empirical analysis} of software aging, e.g., to measure the impact of software aging effects in real systems \citep{garg1998methodologyaging,silva2006soap,grottke2006agingwebserver,carrozza2010memory,araujo2014software} and to get insights on aging-related bugs \citep{grottke2007fightingbugs,machida2012aging,cotroneo2013fault,grottke2010empirical}. We here briefly review the key contributions in the empirical branch, to provide background for the experimental methodology of this study.

\vspace{4pt}
\noindent
\textbf{Software aging metrics and memory consumption.} 
In an early study on software aging in a network of UNIX workstations, \citet{garg1998methodologyaging} collected resource utilization metrics using SNMP, which included memory, swap, filesystem, and process utilization metrics. They found that many of the failures (33\%) were indeed caused by software aging (e.g., resource exhaustion failures), and that \emph{memory} was the resource exhibiting the shortest \emph{time-to-exhaustion} among all the monitored resources. This observation has also been consistently confirmed in subsequent empirical studies, such as in server applications \citep{grottke2006agingwebserver}, middleware \citep{carrozza2010memory}, and cloud management software \citep{araujo2014software}.

\vspace{4pt}
\noindent
\textbf{Trend analysis for software aging.} 
Statistical hypothesis testing (in particular, the \emph{Mann-Kendall test} and the \emph{seasonal Kendall test}) have often been used (e.g., by \citet{garg1998methodologyaging} and \citet{grottke2006agingwebserver}) to point out \emph{degradation trends} in resource utilization, in order to understand whether the time series of resource measurements show a gradual increase/decrease. Moreover, regression techniques, such as the \emph{Sen's procedure} \citep{garg1998methodologyaging} and \emph{autoregressive models} \citep{grottke2006agingwebserver}, have been applied to forecast the time-to-exhaustion of resources. 
One important challenge is represented by the variability workload conditions and configuration, as opposed to laboratory tests in which these factors can be controlled. Thus, subsequent studies have adopted machine learning techniques \citep{alonso2010adaptive}, such as \emph{decision trees} and robust time series analysis techniques \citep{zheng2014automatic}, such as the \emph{Cox-Stuart test} and the \emph{Hodrick-Prescott filter}, to make the aging detection less dependent on assumptions about the types of aging trends (e.g., seasonal or non-linear trends) and about the workload (e.g., by establishing a relationship between workload and aging trends).

\vspace{4pt}
\noindent
\textbf{Stress testing.} 
Stress testing is a potential means to accelerate the onset of software aging phenomena, by providing high volumes of workload (e.g., a high number of client requests) for a long amount of time. For example, this approach has been used by \citet{silva2006soap} and \citet{carrozza2010memory} respectively on a web server, a SOA system, and a CORBA middleware, showing that aging can lead to gradual performance degradation in terms of throughput, latency, and success rate of requests.

\vspace{4pt}
\noindent
\textbf{Aging-related bugs.} 
Empirical research has looked into bugs that have been causing software aging issues in complex software systems. One early study has been conducted on embedded software used for space missions \citep{grottke2010empirical}, which pointed out that \emph{Mandelbugs} (i.e., ``transient'' software faults, which aging-related bugs belong to) are a significant part (higher than $30\%$ on average). This finding has later been confirmed in open-source software projects for the LAMP stack \citep{cotroneo2013fault}, and cloud computing \citep{machida2012aging}. Most of the aging-related bugs affect memory management (e.g., memory leaks). 
Several techniques have been developed to aid software engineers in finding software aging bugs. In particular, most of the research literature on debugging has been focused on memory management bugs. Popular tools for findings \emph{memory leaks} (i.e., objects that have not been deallocated) in C and C++ software include \emph{Valgrind} \citep{nethercote2007valgrind} and \emph{AddressSanitizer} \citep{serebryany2012addresssanitizer}, which traverse at run-time the heap memory, starting from the program's root set of references (e.g., pointers on the stack) and following the pointers inside the objects to find the unreachable (i.e., leaked) ones. For the Java programming language, which avoids memory leaks with automated garbage collection, research has been focused on \emph{memory bloat}, i.e., unnecessary objects that are still referenced by the program but will not be used, thus increasing the load for memory management and impacting on the performance of the program \citep{Ghanavati2019}.  \citet{xu2013precise} and \citet{jump2007cork} analyzed in detail memory bloat issues caused by the accumulation of unused objects in Java containers (e.g., the \emph{HashMap} and \emph{ArrayList} data structures), and proposed techniques for analyzing the heap growth over time to identify objects suspected of bloating memory, based on the age, the time-to-last-access, and the contribution to memory consumption. \citet{sor2013improving} with the Plumbr tool, and \citet{andrzejak2017detection}, used heap utilization metrics in conjunction with machine learning in order to rank the suspicious objects. To further support developers, \citet{xu2011leakchaser} developed \emph{LeakChaser} to allow them to annotate a program with \emph{transactions}, to infer and to check the scope of objects, and with \emph{assertions}, to specify lifetime relationships between objects. Later on, the \emph{PerfBlower} approach by \citet{fang2015perfblower} introduced a domain-specific language, in which developers define \emph{penalty counters} for objects in the program, which are then increased or zeroed at run-time (``amplify'' and ``deamplify'') when some developer-defined conditions occur.

\vspace{4pt}
\noindent
\textbf{Software aging in mobile systems.} 
The initial studies on software aging in mobile devices were focused on \emph{user applications}. \citet{araujo2013investigative} performed stress tests in Android, and pointed out aging trends in popular Android applications, such as Foursquare. Other studies proposed techniques to off-load tasks to the cloud and to perform local application restarts in order to prevent the performance degradation of mobile apps \citep{wu2015software,wang2015reducing}, and to profile and to diagnose performance bugs in UI-triggered asynchronous applications \citep{kang2015persisdroid}.

More recent studies have been investigating software aging and rejuvenation in the Android OS. The empirical study of bug reports by \citet{qin2017empirical} found that the percentage of aging-related bugs and Mandelbugs in the Android OS is comparable to other open-source projects of similar size and complexity. \citet{qiao2018two} and \citet{xiang2018new} presented model-based rejuvenation studies, respectively based on Markov chains and Stochastic Petri Nets, for app-level and OS-level rejuvenation, by taking into account the lifecycle of mobile applications. Other studies by \citet{huo2018using}, \citet{qiao2018empirical}, and \citet{weng2017rejuvenation} have instead developed measurement-based techniques for forecasting resource utilization and performance degradation, by applying machine learning techniques.

\vspace{4pt}


\section{Research questions}
\label{sec:research_questions}

In this study, we consider the following research questions:

\FrameSep4pt

\begin{framed}
\noindent
  \begin{center}
    \textit{(RQ1) Is the Android OS affected by software aging effects? Does software aging affect responsiveness and/or resource utilization, and to what extent?}
  \end{center}
\end{framed}

This is the basis to explore software aging problems in Android, and to motivate further research on mitigating such possible problems through better testing or rejuvenation. To answer this question, we consider a wide set of experimental conditions, in terms of workloads and device configurations, that may influence the extent of software aging effects. Moreover, we measure the user-perceived responsiveness of the Android devices over time, to analyze whether software aging has an impact on the user experience. This research question is further divided into two questions, in order to assess the extent of software aging from different perspectives.

\begin{framed}
\noindent
  \begin{center}
    \textit{(RQ1.1) Is software aging widespread across Android devices from different Android vendors?}
  \end{center}
\end{framed}

This stems from the extensive customizations that Android vendors introduce in their products \citep{iannillo2017chizpurfle}. Therefore, the open-source version of Android (AOSP) is typically extended by vendors with new graphical interfaces and services to enhance the user experience. However, these customizations can also introduce new software defects, and exacerbate software aging issues. Thus, we consider Android devices from four different vendors, so as to figure out whether software aging is a common problem among them and whether vendor customization can worsen software aging issues. 
It is worth noting that the goal of the analysis is not to make claims about which device vendor is most prone to software aging, which would require a different experimental design than our one to cover multiple devices from the same vendor.  Instead, we want to analyze whether the problem of software aging exists across more than one specific vendor, and whether there exists a variability among vendors due to their proprietary customizations.

\begin{framed}
\noindent
  \begin{center}
    \textit{(RQ1.2) Is software aging widespread across different versions of the Android OS?}
  \end{center}
\end{framed}

This is motivated by the extensive architectural revisions that the Android OS has been undergoing in recent years.  The Android OS has been introducing new services, new features for managing security (e.g., stricter permission management mechanisms), new user interfaces (e.g., to improve accessibility), and new mechanisms for managing memory and power utilization. Moreover, vendor customizations also vary across products. Therefore, we compare software aging across different Android versions, in order to investigate whether the evolution of Android had any influence on software aging issues.

\begin{framed}
\noindent
  \begin{center}
    \textit{(RQ2) Where are the software aging problems that affect the Android OS located?}
  \end{center}
\end{framed}

In order to better understand software aging phenomena, we investigate resource utilization metrics at process- and at task-level, by correlating them with the user-perceived performance degradation. This information is useful to identify where to put more efforts  in the Android OS in order to either prevent software aging (e.g., by more extensive tests and debugging) or to mitigate it through software rejuvenation (e.g., by cleaning-up or restarting selected components, rather than the whole Android device).

\section{Experimental methodology}
\label{sec:methodology}

To analyze software aging issues in Android, we adopt an experimental methodology based on \emph{stress testing}. 
In this paper, the term ``test'' is used to denote an experiment where the system is exercised with an intensive workload for a long period (typically, several hours), in order to increase the likelihood that software aging effects, such as memory leaks, accumulate over time \citep{grottke2006agingwebserver,silva2006soap,carrozza2010memory}. The test execution ``fails'' when it exhibits software aging anomalies that lead to poor responsiveness and crash/hang of the system. Test results are then analyzed to identify which conditions exacerbate the occurrence of these failures.

We perform tests under several different conditions, as the extent of software aging effects (\emph{e.g., }the rate at which the device experiences performance degradation or resource depletion) varies depending on how the system is configured and exercised \citep{carrozza2010memory}. For example, in the context of Android OS, different user apps may have a different impact on software aging, as they may trigger different OS services; or different OS configurations (e.g., vendors or versions) may exhibit different software aging effects. However, considering all of the possible combinations of workloads and configurations leads to an extremely high number of long-running experiments, which would take an unfeasible amount of time to complete.

Thus, in order to determine the most influential factors  (\emph{e.g.,} workload, device vendor, OS version, etc.) on the software aging trend, we design a test plan for the Android OS.
Specifically, we define a set of \emph{factors} (\emph{i.e.,} the parameters of a test) and their possible values (also called \emph{levels}  \citep{montgomery2008design}). First, we identify the feasible combinations of Android devices and Android versions, since we can not install all the versions in all the devices. Then, we define an experimental design by investigating different combinations of these factors, according to the research questions. We introduce \emph{response variables} to quantify the impact of a test on the target device in terms of software aging, and correlate the factors with the response variable to identify the most influential ones. We consider both \emph{user-perceived} response variables (\S{}~\ref{subsec:user_perceived}), and \emph{system-related} response variables (\S{}~\ref{subsec:system_perceived}). 

The first objective of the data analysis is to assess whether the software aging is present in Android OS, by analyzing the \emph{user-perceived} response variables (\cfr RQ1). Once we demonstrate the presence of software aging, we analyze the variance of these response variables with respect to the values of the factors. In particular, we focus on the variances between different Android vendors (\cfr RQ1.1) and between different versions of Android OS (\cfr RQ1.2). Finally, we analyzed the \emph{system-related} response variables to unveil the underlying components where the software aging phenomenon is internally localized (\cfr RQ2).

\subsection{Statistical Analysis}
We analyze the experimental data using statistical techniques (further discussed in the next subsections), including:

\begin{itemize}

\item The Mann-Kendall (MK) test to statistically assess if there is a monotonic trend in a series of the variable of interest over time \citep{garg1998methodologyaging}. 
Although it is the most used tests for aging analysis \citep{cotroneo2014survey}, it may be subject, like any other hypothesis test, to false positives/negatives \citep{Machida2013}: in order to double-check the results, we apply the following further tests for trend detection \citep{onoz2003power}. Their effectiveness depends on the shape of the data distribution: \textit{i)} the Cox-Stuart test, \textit{ii)} the \textit{t-}test, and \textit{iii)} the Spearman's rho test. We claim that there is a trend only if the MK test (which is the most used ones for aging analysis) plus two further tests (out of the three mentioned above) reject the null hypothesis of no trend.

\item The Sen's procedure to compute, in a non-parametric way, the slope of a trend \citep{sen1968estimates,theil1992rank};

\item The Spearman's rank correlation coefficient to analyze the statistical dependence between two variables of interest  \citep{pirie1988spearman}.

\item The Analysis of Variance (or ANOVA) and the Kruskal-Wallis/Wilcoxon hypothesis test to analyze whether the differences among two sets of experiments are \emph{statistically-significant} (\emph{i.e.,} not simply due to random variations) \citep{anscombe1948validity,daniel1990kruskal}.

\end{itemize}

Under fixed experimental conditions, which reflects our case, the system is expected to exhibit stable performance throughout the experiment: we apply a fixed workload, in which events are generated at a constant rate, and exercise the applications in the same way over and over, where the apps are periodically killed and their resources are freed-up. The experiments have a long duration and the trends are identified through statical trend analysis in order to account for noise and transient fluctuation.

\subsection{User-Perceived Response Variable}
\label{subsec:user_perceived}

To quantify software aging as perceived by users, we focus on the \emph{responsiveness} of the Android OS, as it is a key design goal of this mobile OS. 
In particular, we analyze the \textbf{Launch Time} (LT) of Android Activities (\emph{i.e.,} an application component that provides a GUI screen), which is the period between the request to start an Activity, and the appearance of the Activity on the screen, including the initialization of background and foreground elements. The LT  measures the end-to-end latency between the action of a user (starting a new Activity) and the response of the Android OS, across the whole Android software stack (\ie from the kernel at the lowest level, to the user interface at the highest level). This includes the update of the UI, the initialization of the Android run-time, the allocation of new processes and threads, the registration of a new Activity by the Activity Manager, permission checks by the Package Manager, \emph{etc.}. Indeed, the Android project adopted this metric as an early design goal (``cold-start a basic application, up to a responsive GUI, within $200 ms$ at most'' \citep[ch. 10.8]{tanenbaum2014modern}, \citep{android-web-perf-anr}, and clearly correlates the user expectation with the Activities launch times \citep{android-app-start-time}. 
We start our analysis by focusing on the LT metric (rather than resource utilization metrics, that are discussed in the next subsection) since this metric directly impacts on the user, i.e., it is the key performance indicator for the ``quality of experience'' from the perspective of the users. Resource consumption metrics (memory, CPU) may also exhibit aging trends, but these metrics have an indirect relationship with performance and failures, and have relevance only if there is some user-perceived performance degradation or failure (e.g., if a memory consumption trend does not affect responsiveness or cause failures, then it may be negligible for users and vendors). Thus, we analyze resource utilization metrics in a later stage, to get more insights about responsiveness degradation and the components and causes of software aging (\cfr~\ref{subsec:system_perceived}).

We measure the LT by analyzing the logs from the \emph{Activity Manager} of the Android OS, which is the service responsible for instantiating new activities and to switch among them by saving and restoring their state.  We collect these logs using the Android \emph{logcat} utility. 
As an example, the following line is the log message that shows the ``MainActivity'' Activity from the application ``com.example.myapp'', which took \emph{100 ms} to complete its initialization:

\begin{lstlisting}[frame=single, basicstyle=\footnotesize]
I/ActivityManager(1097): Displayed com.example.myapp
    /.MainActivity: +100ms
\end{lstlisting}

During an experiment, if an Activity is retrieved from a cache  when the user switches an app, the LT cannot be measured. For this reason, periodical samples of LT are obtained by periodically terminating and restarting workload applications, with a frequency of 1 minute. This is necessary for preventing the OS to cache the Activities.

Moreover, by periodically restarting the apps, we avoid that software aging effects (such as leaked memory) could accumulate inside the apps since our focus is not to study aging of Android apps, but rather the software aging effects in the underlying Android OS. 

For each experiment, we analyze the LT to identify any degradation of responsiveness. To this aim, we produce a time series for each experiment using the LT samples of all activities collected during the experiment, and we apply the non-parametric \emph{Mann-Kendall} (MK) statistical test to check whether the time series exhibits a \emph{trend} \citep{garg1998methodologyaging}, along with the three further trend tests mentioned previously.  
All of them check the null hypothesis that there is no monotonic trend in the time series, and provides a level of significance (p-value)  for the likelihood that the null hypothesis is actually true in the time series. If the p-value is lower than a given $\alpha$, we can reject the null hypothesis with probability (namely \textit{confidence}) greater than $(1-\alpha)$, which points out that the LT has been affected by a trend. We require that the confidence should be higher than 95\% ($\alpha=0.05$) for the MK test and for at least two further tests out of the three ones mentioned above. 

We mainly rely on the MK test, since it is a non-parametric test, it does not require the measurements to follow a specified distribution or the trend to be linear. However, it is required that measurements are not serially correlated over time. To apply the test, we preliminarily checked for auto-correlation of each data series by means of the \textit{Durbin Watson} test \citep{durbin1951testing} at $\alpha=0.05$: when the test statistic $d$ and the value $(4-d)$ are greater than the upper critical value, then there is no evidence of positive or negative auto-correlation: in such a case the conventional MK test is applied. In the other cases (\emph{i.e.,} there is evidence of auto-correlation or the test is inconclusive) a modified version of the MK test is adopted, namely the \textit{Hamed and Rao Variance Correction Approach} \citep{hamed1998modified}. The approach removes the trend from the series and computes the effective sample size significant serial correlation coefficients. A corrected p-value is then provided as an outcome.

If the (modified) MK test indicates the presence of a  trend in the LT, we then obtain the \emph{slope} of such trend by applying the \emph{Sen's procedure} \citep{sen1968estimates,theil1992rank}, which is a non-parametric, robust technique that fits a linear model and computes the rate at which the samples increase over time. It simply computes the slope as the median of all slopes between paired values, and it is insensitive to outliers. This approach is often adopted in software aging studies where the system is stressed under fixed conditions, which is likely to lead to a fixed degradation rate, if any \citep{grottke2006agingwebserver,silva2006soap,carrozza2010memory}.
It should be noted that the outcome of each experiment is not merely based on a single-point observation, in which case there is no clue of the variability of the error caused by possibly repeating the run. It is instead a \textit{trend} computed over hundreds of observations, and its slope value, obtained by the Theil-Sen estimator on the data series of ``response times'', is associated with a confidence interval under confidence level of 95\% -- thus its assessment accounts for the impact of the variability of response times. This is a good compromise between repeating more times the same 6-hours-long run for each of the 72 experiments (which would be more accurate, but too costly), or having only 72 single-point observations with no statistical significance associated with the response variable. The detailed experimental plan is presented in Subsection~\ref{subsec:exp_plan}.

\subsection{System-Related Response Variables}
\label{subsec:system_perceived}
In our analysis, system-related metrics are important to point out which are the most stressed areas of the Android OS that might be causing software aging.
System-related metrics include the \emph{memory utilization} (which is the resource most exposed to software aging issue due to memory management bugs; and a scarce one for mobile devices); the \emph{CPU utilization} (which is also exposed to software aging, e.g., due to algorithmic bugs that waste CPU time on bloated data structures); and the \emph{garbage collection duration} (which is a critical activity for the efficient use of memory). 


\subsubsection{\textbf{Memory}}

Many previous experiments on software aging effects indicated that this resource is the most affected one and it tends to have the shortest time-to-exhaustion (TTE) \citep{garg1998methodologyaging,grottke2006agingwebserver,silva2006soap,carrozza2010memory,cotroneo2013fault}. The Android OS uses complex mechanisms to manage memory, by automatically handling the lifecycle of apps (e.g., collecting resources once an app is not used for a long time, e.g., through the \emph{Low Memory Killer} (LMK) mechanism \citep{tanenbaum2014modern}), by recycling processes (e.g., when starting a new Activity), and by managing memory inside applications based on the ART (\emph{Android Run-Time}) Java environment. Another potential cause of aging effects in memory utilization is represented by the complexity of the Android OS services, such as \emph{Activity Manager} and \emph{Package Manager}, that are persistent and may accumulate aging effects over time.


We analyze memory utilization through the Android \emph{dumpsys} utility, which reports the memory consumption of the Android OS both in user-space (e.g., the memory used by Android apps and services) and in kernel-space (e.g., Android extensions to the Linux kernel such as the Kernel Samepage Merging, \emph{KSM}, and virtual memory compression, \emph{zram}). We analyze memory consumption of each process of the Android OS, by periodically launching the command \emph{dumpsys} every 30 seconds and extracting the \emph{Proportional Set Size} (\emph{PSS}) of the process, \emph{i.e.,} the footprint of the process on the physical RAM (\emph{e.g.,} not including parts of the process that do not consume memory, such as program code that has not been executed and that still resides on the storage). We focused on this metric because our previous results show that it is strongly correlated to performance degradation trends \citep{cotroneo2016software}.

%
We check again whether LT degradation is related to per-process PSS metrics. For each PSS series, we perform the following two steps: (i) we test the presence of a trend (and compute its slope) using the four trend tests, i.e., the (modified) Mann-Kendall, Cox-Stuart, \textit{t-test}, Spearman's rho tests, and the Sen's procedure; (ii) we compute a \emph{correlation measure} between the slopes of the metric and the slopes of the median LT trend, across all experiments, using the non-parametric \emph{Spearman's rank correlation coefficient} \citep{pirie1988spearman}, since it is robust to outliers and does not make restrictive assumptions on data, contrarily to the parametric counterparts. The correlation points out whether a trend of the metric is systematically accompanied by a degradation trend of the LT. 

\subsubsection{\textbf{Garbage Collection}}
\label{subsec:garbage_collection}

Garbage collection (GC) is a key component of the modern programming environment, as it manages dynamic memory allocations in place of the programmer (\emph{e.g., }freeing unused area) in order to avoid memory management bugs. However, despite garbage collection, there can still be residual software aging effects: if unused objects are still referenced by the program (\emph{e.g.,} due to poor object handling by programmers), the GC is not able to dispose of the objects, which can accumulate over time \citep{carrozza2010memory,STVR2013}. 
This problem is exacerbated by memory fragmentation, which occurs when the program allocates a mix of small and large objects with different lifetimes, causing ``holes'' in the heap area and increasing the duration of garbage collections and of memory allocations for large objects; and by other bad memory management practices (\emph{e.g.,} frequently re-allocating objects that could instead be reused). 
Because of these phenomena, GC can significantly degrade the performance perceived by users.

In our experiments, where the device executes under fixed conditions, the GC collection times are expected to be stable throughout the experiment, and the presence of a trend definitely shows an anomaly. In particular, such trends can occur when numerous and scattered objects accumulate in the heap, which the GC will need to walk through object references to reclaim memory. Therefore, the increasing GC time is likely due to the growth and fragmentation of the heap memory. If GC takes more and more time, a process can be noticeably slowed down during GC. 
This phenomenon has been found in several studies on software aging, which showed that performance degradation in real systems was caused by the higher memory management overhead, which in turn was caused by memory bloat \citep{carrozza2010memory,grottke2006agingwebserver}. 

For the above reasons, we include the duration of GC among system-related metrics.  
We collect information on GC from the logs of the Android OS, marked with the ``art'' tag. The Android RunTime (ART) is the Android equivalent for the JVM. The ART reports on GC only in the case that the GC takes much more than usual (in particular, when the \emph{GC Pause Time} exceeds 5 ms or the \emph{GC Duration} exceeds 100 ms). In such a case, the log includes the event that triggered the GC, the GC algorithm, the amount of time spent for the GC, the number of objects freed by the GC, and the available heap memory. 
We collect these logs as soon as they appear during the experiments. 
We expect that intensive workloads (such as the ones used by our stress tests) can highlight the effects of poor memory management in Android components, which in turn can result in degraded performance.

The main source of information on GC is the logs provided by the Android Runtime, which can be queried using the \emph{logcat} tool. GC logs are denoted by the ``art'' tag, as in the following example:

\begin{lstlisting}[frame=single, basicstyle=\footnotesize]
I/art: Explicit concurrent mark sweep GC freed 
    104710(7MB) AllocSpace objects, 21(416KB) 
    LOS objects, 33%free, 25MB/38MB, paused 
    1.230ms total 67.216ms
\end{lstlisting}

%
The GC metrics are analyzed for each process, by computing trends using the (modified) Mann-Kendall test and the Sen's procedure. We count the number of cases in which the process exhibited an increase of GC times, which reveals a potential relationship between software aging (in particular, loss of responsiveness) and memory bloat or fragmentation.

\subsubsection{\textbf{CPU and memory utilization at the task level}}
\label{subsec:task_level_metrics}

The Android OS adopts a complex multi-process and multi-threaded architecture to run its several services and components (e.g., to manage a specific hardware resource or provide an API). Since the previous metrics do not provide specific information about individual threads inside a process, we introduce additional metrics to get more insights into the activity of individual services running inside threads.

In the context of the Linux kernel, both processes and threads are internally represented by a \emph{task\_struct} object (we use the term \emph{task} in the following of this paper), where a thread is a special type of task that shares certain resources with other tasks (i.e., their \emph{task\_struct}s point to the same file table, the same page table, etc.) \citep{love2010linux}.
Therefore, we analyze CPU and memory utilization metrics for individual tasks. These metrics point out which tasks are mostly active during the onset of software aging effects, and are a potential origin of software aging.


We obtain task-level metrics from the \emph{proc} filesystem of the Linux kernel. In particular, we use the virtual files \emph{schedstat} and \emph{stat} files that are exposed by the Linux kernel (in the virtual folder /proc/TASK\_PID/) to provide information on CPU and memory activity of tasks. These metrics include the number of \emph{major page faults} (i.e., memory accesses that trigger storage accesses to retrieve the data) and \emph{minor page faults} (i.e., memory accesses that are served by re-using data from the storage cache or shared from other processes), and CPU execution time spent in user-space and kernel-space, which respectively denote the CPU and I/O activity of the task. We periodically sample these task-level metrics every 30 seconds.

To identify critical tasks, we compute trends for each metric and for each task using the (modified) Mann-Kendall test and the Sen's procedure. Then, we count the number of cases in which a metric exhibited a statistically-significant trend for the task, at a confidence level of 95\%. The higher the count, the higher the likelihood that the metric evolves with software aging effects, thus revealing a potential relationship between a task and software aging of the device.

\subsection{Factors and Levels}
\label{subsec:factors}

We consider several \emph{factors} to cover different configurations and workloads in the experimental plan. In particular, we define 5 factors and obtain the test plan combining the \emph{levels} of these factors. Factors and levels are summarized in \tablename{}~\ref{table:factors_and_levels}.

\vspace{5pt}
\noindent
$\triangleright$
\textbf{\textit{Device (DEV).}}
The Android devices in our experimental setup represent the levels for the DEV factor. 
We conducted experiments on high-end smartphones from four different, popular vendors. 
Although market data is highly variable and difficult to gauge, there is a consensus among market analysts that Samsung and Huawei are the two most popular Android vendors, and that their \emph{S} and \emph{P} series were sold at high volumes; for example, see the analysis by the Counterpoint market intelligence firm \citep{counterpoint-market}. LG and HTC have also been very popular brands. Both of them have been producers of smartphones for Google devices, including the Nexus LG analyzed in this work, and more recently HTC for the Pixel series. The Nexus device is an interesting target for experimentation, since these devices are designed by Google to be ``reference'' devices for the Android platform, with more timely updates from Google and more lightweight vendor customizations \citep{wikipedia-nexus}. Moreover, the HTC One series has been quite popular until recently with the rise of the competition, and these devices have been designed to be top-quality Android products \citep{android-central-htc}. 
In total, we have four levels for the DEV factor, labeled as \emph{HTCONEM9} (\emph{HTC M9 One}), \emph{HUAWEIP8} (\emph{Huawei P8}), \emph{LGNEXUS} (\emph{LG Nexus}), and \emph{SAMSUNGS6EDGE} (\emph{Samsung S6 Edge}).

\vspace{5pt}
\noindent
$\triangleright$
\textbf{\textit{Version (VER).}}
The Android OS versions available for a device determine the levels for the \emph{VER} factor. Given the devices that we selected for experimentation, we have three levels for the \emph{VER} factor: \emph{ANDROID5}, \emph{ANDROID6}, and \emph{ANDROID7}.

\begin{figure}[!t]
\centering
\includegraphics[width=0.8\textwidth]{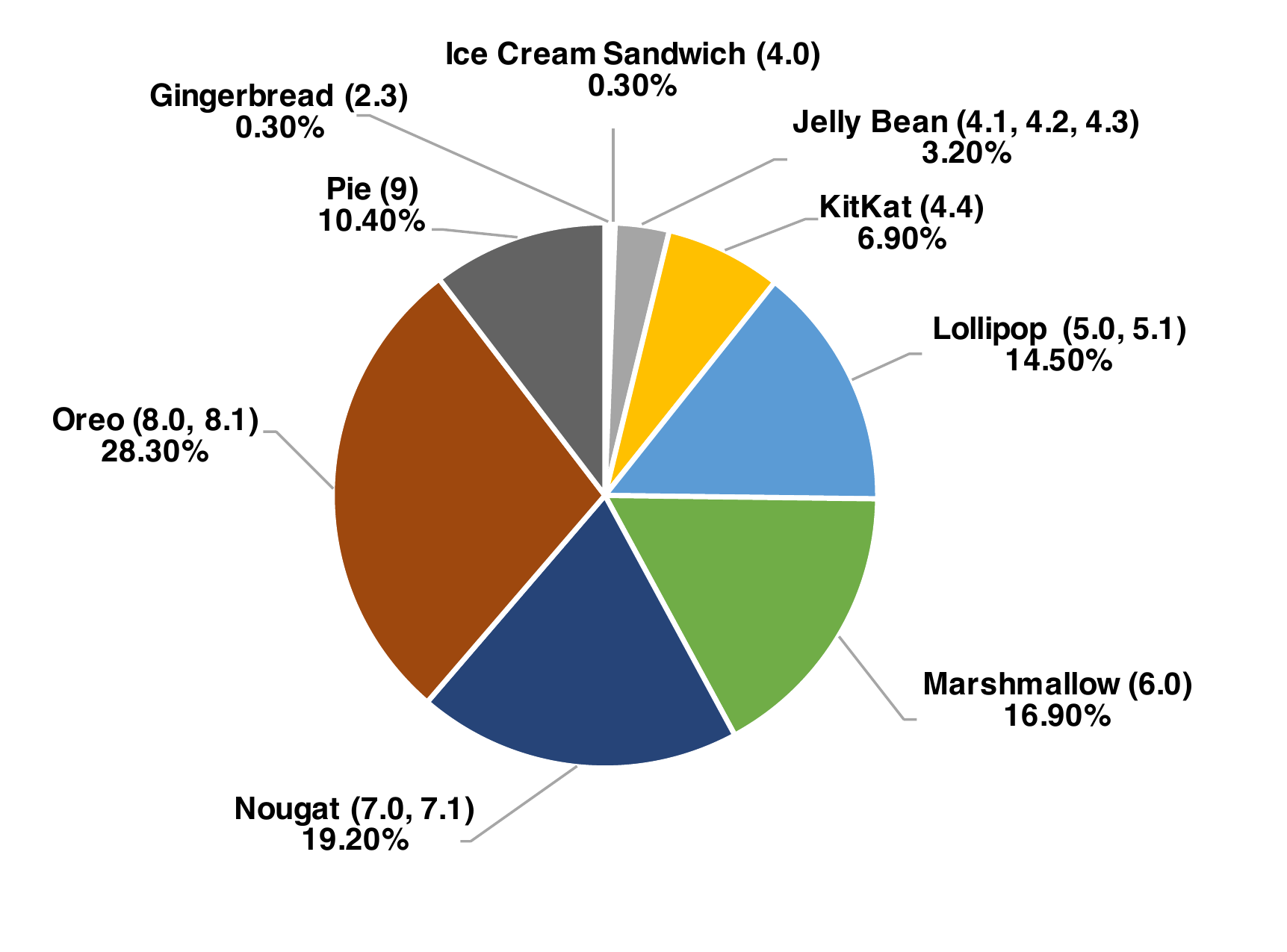}
\caption{Distribution of Android OS versions across users (source: \url{https://developer.android.com/about/dashboards} \citep{android-dashboard}, May 2019).}
\label{fig:android_versions}
\end{figure}

According to recent statistics \citep{android-dashboard}, these versions of the Android OS account for more than 50\% of Android devices on the market up to May 2019 (\figurename{}~\ref{fig:android_versions}). Even if newer releases are available on the market, it is important to note that the adoption of new versions by users has always been quite slow, due to the reluctance of device vendors to provide Android OS updates for a long period, and it is typical for Android devices to receive updates only for just 2 years or even less. In turn, this is caused by the relatively high costs to align a new version of the Android OS with the proprietary customizations by the vendors, and the need for vendors to incentivize users to buy new Android devices. 
We focused our efforts on versions up to 7 as the Android OS reached a high degree of maturity with this version, after 10 years since the initial version. Therefore, even if the Android OS has continued to be developed so far, there have been no major revisions of the fundamental OS architecture compared to past versions (e.g., no major component has been rewritten from scratch as it happened between for the Dalvik VM that was used by Android 4 and lower versions).

It is worth noting that not every level in the \emph{DEV} factor can be combined with every level in the \emph{VER} factor. In our experiments, we only use official Android releases, by flashing factory images publicly released by the device vendors (i.e., the binary blob of the OS to be written on storage memory). Since the vendors only release updates for a limited period of time, not all of the versions 5, 6, and 7 were available for all devices. To force a specific version of Android, we would have needed to install the open-source Android, but that would have not reflected the actual Android running on real devices. If we used different devices to compare the different Android versions, we would also vary the underlying hardware, thus adding uncertainty on whether the differences are due to the different hardware. Despite the lack of some combinations of versions/devices, the available versions were still enough to allow us to compare multiple devices running the same Android version, and to evaluate more than one Android versions on the same device (see Section~\ref{subsec:exp_plan}).

\vspace{5pt}
\noindent
$\triangleright$
\textbf{\textit{Application Set (APP).}}
In our experiments, we use different sets of applications as the workload to exercise the Android OS. These apps are selected to be representative of typical usage scenarios (including browsing, making photos, dialing, chatting), and include both stock apps and third-party apps. We include popular Android applications, listed in \tablename{}~\ref{table:factors_and_levels}. 
We organized applications in two groups, which represent the two levels of the APP factor: European applications (\emph{EU}), and Chinese applications (\emph{CHINA}), which are obtained respectively from the European version of the Google app market, and from Chinese app markets. \footnote{For those applications requiring the user to login,  we have manually configured them before the experiment, by signing up and activating the automatic login.}

\vspace{5pt}
\noindent
$\triangleright$
\textbf{\textit{Workload Events (EVENTS).}}
Our workload generator produces a set of UI events to interact with the apps and the Android device. The events include application switch, touch, motion, trackball, and navigation events. 
This approach is similar to tools for testing UIs in Android, such as \emph{AndroidRipper} \citep{amalfitano2012using}, \emph{Dynodroid} \citep{machiry2013dynodroid}, and \emph{Sapienz} \citep{mao2016sapienz}. Our goal is not to find bugs in a specific app, but to stimulate a collection of apps in order to stress the Android OS at managing the apps. We adopt the popular Google's Android \emph{Monkey} tool \citep{monkey}, which is integrated with the Android OS. 
The events are generated randomly, and their probability of occurrence is configured by us. The EVENTS factor varies across three levels: \emph{MIXED1}, where half of the events are application switches; \emph{MIXED2}, where half of the events are touches; \emph{MIXED3}, where half of the events are navigation events. In every level, the other half of the events are of the remaining types and are selected according to a uniform random distribution. 
The events are generated at a high frequency ($500 ms$) to stress the device and to trigger aging phenomena within the duration of the experiment (6 hours). Setting higher values (lower frequency) would have prolonged the duration of the experiments, which would have required an excessive experimental cost. Instead, setting a lower value (higher frequency) would have caused too many events that the devices could not be able to handle, since the creation of new activities often can take several hundreds of milliseconds.

\vspace{5pt}
\noindent
$\triangleright$
\textbf{\textit{Storage Space Usage (STO).}}
We execute experiments either with or without available storage (\emph{i.e.,} free space for storing data), as this aspect can impact on some of the services of the Android OS (\emph{e.g.,} by storing photos and videos from the camera). This factor varies between two levels: \emph{FULL}, where 90\% of the storage is occupied by filling it with videos and images; and \emph{NORMAL}, where the default amount of storage space is used (\emph{i.e.,} the storage is occupied only by system files and application packages).

\vspace{5pt}

We adopt the one-way \emph{Analysis of Variance} (\emph{ANOVA}) technique \citep{anscombe1948validity} to assess which factors impact the response variable in a statistically-significant way. 
To figure out if (and which type of) ANOVA can be applied, we verified the following assumptions: \textit{i)} independence of treatments; \textit{ii)} normality of residuals; \textit{iii)} homoscedasticity of variances. 
Independence is assured by resetting the environment at each treatment execution and avoiding human biases by making the entire procedure automatic. 
As for normality, we apply the \textit{Shapiro-Wilk} test; the null hypothesis states that data came from a normal distribution. Rejecting the hypothesis means that the assumption is not verified. To verify homoscedasticity, we performed the Levene's test. The null hypothesis is that variances of levels of variables are homogeneous; rejecting the hypothesis means that we are dealing with unequal variances. We then apply the following ANOVA tests accordingly: \textit{i)} if the residuals are normally distributed and homoscedastic, then the conventional parametric ANOVA Fisher test is used;  \textit{ii)} if the residuals are normally distributed but heteroscedastic, then the Welch ANOVA test is used, which admits unequal variances; \textit{iii)} if residuals are not normally distributed, then the non-parametric Kruskal-Wallis/Wilcoxon test is adopted \citep{daniel1990kruskal}.

In all the cases, the null hypothesis is that the factor does not impact the response variable, and the p-value indicates again the confidence in rejecting this hypothesis. We conclude that a factor impacts the response variable if the level of confidence is higher than 95\%, i.e., the p-value is less than 0.05.

\begin{table}[!t]
\caption{Factors and levels}
\label{table:factors_and_levels}
\centering
\fontsize{8}{9}\selectfont
\begin{tabular}{C{1.5cm}C{2.7cm}C{6cm}}
\toprule
\textbf{Factor} & \textbf{Level} & \textbf{Description}\\
\midrule
\multirow{4}{*}{DEV} & HTCONEM9 & HTC One M9 device\\
 & HUAWEIP8 & Huawei P8 device\\
  & LGNEXUS & LG Nexus device \\
  & SAMSUNGS6EDGE & Samsung S6 Edge device \\
\midrule
   \multirow{3}{*}{VER} & ANDROID5 & Android 5 (Lollipop)\\ 
    & ANDROID6 & Android 6 (Marshmallow)\\ 
     & ANDROID7 & Android 7 (Nougat) \\
 \midrule
     \multirow{2}{*}{APP} & EU & com.google.android.videos, com.*.camera, com.android.browser, com.android.email, com.android.contacts, com.google.android.apps.maps, com.android.chrome, com.google.android.play.games, com.android.calendar, com.google.android.music, com.google.android.youtube\\ \cmidrule{2-3}
      & CHINA & com.tencent.mm, com.sina.weibo, com.qiyi.video, com.youku.phone, com.taobao.taobao, com.tencent.mobileqqi, com.baidu.searchbox, com.baidu.BaiduMap, com.UCMobile, com.moji.mjweather \\
\midrule
      \multirow{3}{*}{EVENTS} & MIXED1 & mostly switch events\\ 
       & MIXED2 & mostly touch events\\ 
         & MIXED3 & mostly navigation events \\ 
           \midrule
           \multirow{2}{*}{STO} & FULL & 90\% of storage space usage \\
            & NORMAL & default storage space usage\\ 
            \bottomrule
            \end{tabular}

\end{table}

\subsection{Experimental plan}
\label{subsec:exp_plan}

We defined an ad-hoc experimental plan by identifying different combinations of the factors to answer our research questions. The experimental plan includes 72 experiments, for a total of 18 days of experimentation. The experiments are listed in \tablename{}~\ref{table:rq_exp_map} (grouped by research question) and in \tablename{}~\ref{table:experimental_plan} (sorted from the oldest to the newest Android version).

The experimental plan contains three overlapping sets, where each set can keep out one factor and combines the remaining factors in a full-factorial design. 
The first set (EXP13$\sim$EXP60) covers all of the \emph{DEV} levels, and locks the \emph{VER} factor to \emph{ANDROID6}, since Android 6 Marshmallow is the only version that can be installed on all the devices, allowing us to study the impact of software aging across devices from different vendors (and all other factors with the same level). The second set of experiments (EXP1$\sim$EXP24) fixes the \emph{DEV} factor to \emph{HUAWEIP8}, and varies the \emph{VER} factor between \emph{ANDROID5} and \emph{ANDROID6}. The third set (EXP49$\sim$EXP72), instead, fixes \emph{DEV} to \emph{SAMSUNGS6EDGE} and the \emph{VER} to either \emph{ANDROID6} or \emph{ANDROID7}. These last two sets are used to study the impact of software aging across different versions of the Android OS. In each set, with the only exception of the fixed factor, we consider every possible combination of the levels.

\begin{table}
\caption{Experiments Analyzed to Answer the Research Questions}
\label{table:rq_exp_map}
\centering
\begin{tabular}{C{8.5cm}C{2.5cm}@{}m{0pt}@{}}
\toprule
RESEARCH QUESTIONS & EXPERIMENTS &\\
\midrule
(RQ1) Is the Android OS affected by software aging effects? Does software aging affect responsiveness and/or resource utilization, and to what extent? & EXP1$\sim$EXP72 &\\
\midrule
(RQ1.1) Is software aging widespread across Android devices from different Android vendors? & EXP13$\sim$EXP60 &\\
\midrule
(RQ1.2) Is software aging widespread across different versions of the Android OS? & EXP1$\sim$EXP24, EXP49$\sim$EXP72 &\\
\midrule
(RQ2) Where are the software aging problems that affect the Android OS located? & EXP1$\sim$EXP72 &\\
\bottomrule
\end{tabular}
\end{table}

Due to the long duration of the experiments, we adopted an experimental design without replications, namely one long-running test for each of the 72 experiments. 
It should be noted that the ``response variable'' of our experimental design, namely the outcome of a single run, is not a point-value of the metric of interest (e.g., a value of launch time, or memory PSS), but it is a \textit{trend value} (e.g., ``launch time trend''), obtained after a long-running test by the Sen's estimator on the data series of ``launch times'', and is itself associated with a confidence interval under confidence level of 95\%. The fluctuations of the metric under analysis (e.g., launch time) within a single run are  accounted for, from the statistical point of view, by the trend detection test. When we reject the null hypothesis of ``no trend in data'' with 95\% of confidence, we are saying that the fluctuations of the metric under analysis (e.g., the variance of the metric in that single repetition) are not enough large to make us accept the null hypothesis. 
The repetitions would allow having a number of \textit{trend} values for that configuration (since repeating the run would provide a different trend), thus accounting for possible fluctuation of the trend (namely, the variance of the trend values across multiple repetitions). 
We compensate for the replications by applying the Sen's procedure to estimate the slope of the trend. The procedure provides a probability distribution for the slope, and we consider the median value: this value represents the likely value that would be obtained if we repeat the test multiple times. The longer the time series is when applying the Sen's procedure, the more reliable is the resulting trend, with a tighter confidence interval, hence the more likely a new repetition will yield a similar trend.

Thus, we calibrated the duration of a test so as to obtain very stable trends (namely tight confidence intervals) as follows. 
We have analyzed the experimental data from our initial work on software aging in Android \citep{cotroneo2016software}. In that analysis, we conservatively set the duration of tests to 20 hours, since this duration sufficed to make software aging to surface in other systems analyzed by previous experimental studies \citep{carrozza2010memory,bovenzi2011workload}. It turned out that, in the cases where we found aging trends, the trend was apparent much before the end of the 20 hours, as 6 hours of data would have been sufficient to detect all trends, and to estimate the trend slope with less than 1\% error. Therefore, since in this work we are performing a much larger number of experiments than the past work \citep{cotroneo2016software}, we re-calibrate the duration of tests for this work to 6 hours.

When we investigate each research question (\emph{e.g.,} the impact of the Android OS version), we split our dataset into two partitions, where each partition has the same size with the same experimental configurations, but differs pairwise with respect to the investigated factor. For example, the configuration with device=LGNEXUS, workload=MIXED1, storage=NORMAL, and app=EU, appears in both partitions, differing only with respect to OS versions. Under the null hypothesis that the factor does not influence software aging, the results in the two subsets of experiments should not be significantly different. Otherwise, we report that the factor had a statistically-significant impact.

\begin{figure}
    \centering
    \includegraphics[width=0.8\textwidth]{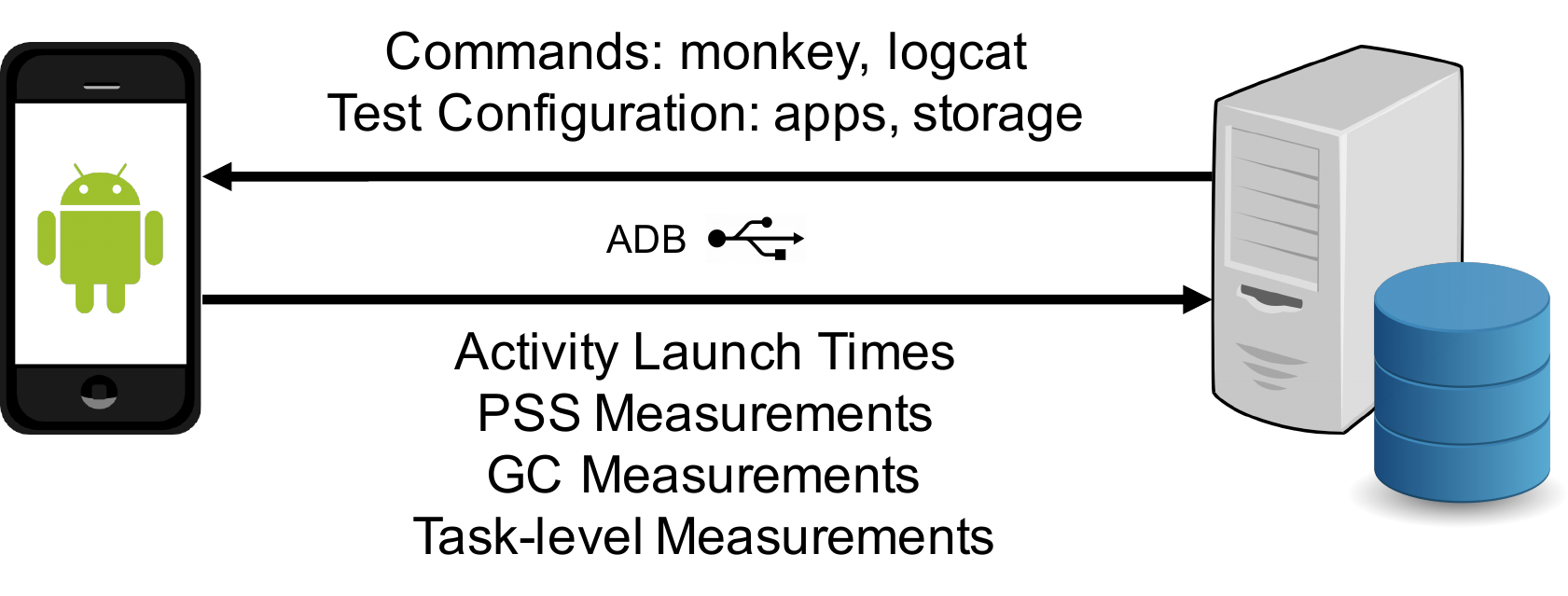}
    \caption{The Experimental Android Testbed}
    \label{fig:testbed}
\end{figure}

The devices are controlled and monitored using the \emph{adb (Android Debug Bridge)} utility, which is a non-intrusive, dedicated channel through the USB port for debugging purposes. The experimental testbed is shown in \figurename{}~\ref{fig:testbed}.


\begin{table}[]
\caption{Experimental plan of the case study}
\label{table:experimental_plan}
\centering
\fontsize{6}{7}\selectfont
\begin{tabular}{cccccc}
\toprule
ID & DEV & VER & APP & EVENTS & STO\\
\midrule
EXP1 & HUAWEIP8 & ANDROID5 & EU & MIXED1 & NORMAL \\
EXP2 & HUAWEIP8 & ANDROID5 & EU & MIXED1 & FULL \\
EXP3 & HUAWEIP8 & ANDROID5 & EU & MIXED2 & NORMAL \\
EXP4 & HUAWEIP8 & ANDROID5 & EU & MIXED2 & FULL \\
EXP5 & HUAWEIP8 & ANDROID5 & EU & MIXED3 & NORMAL \\
EXP6 & HUAWEIP8 & ANDROID5 & EU & MIXED3 & FULL \\
EXP7 & HUAWEIP8 & ANDROID5 & CHINA & MIXED1 & NORMAL \\
EXP8 & HUAWEIP8 & ANDROID5 & CHINA & MIXED1 & FULL \\
EXP9 & HUAWEIP8 & ANDROID5 & CHINA & MIXED2 & NORMAL \\
EXP10 & HUAWEIP8 & ANDROID5 & CHINA & MIXED2 & FULL \\
EXP11 & HUAWEIP8 & ANDROID5 & CHINA & MIXED3 & NORMAL \\
EXP12 & HUAWEIP8 & ANDROID5 & CHINA & MIXED3 & FULL \\
\midrule 
EXP13 & HUAWEIP8 & ANDROID6 & EU & MIXED1 & NORMAL \\
EXP14 & HUAWEIP8 & ANDROID6 & EU & MIXED1 & FULL \\
EXP15 & HUAWEIP8 & ANDROID6 & EU & MIXED2 & NORMAL \\
EXP16 & HUAWEIP8 & ANDROID6 & EU & MIXED2 & FULL \\
EXP17 & HUAWEIP8 & ANDROID6 & EU & MIXED3 & NORMAL \\
EXP18 & HUAWEIP8 & ANDROID6 & EU & MIXED3 & FULL \\
EXP19 & HUAWEIP8 & ANDROID6 & CHINA & MIXED1 & NORMAL \\
EXP20 & HUAWEIP8 & ANDROID6 & CHINA & MIXED1 & FULL \\
EXP21 & HUAWEIP8 & ANDROID6 & CHINA & MIXED2 & NORMAL \\
EXP22 & HUAWEIP8 & ANDROID6 & CHINA & MIXED2 & FULL \\
EXP23 & HUAWEIP8 & ANDROID6 & CHINA & MIXED3 & NORMAL \\
EXP24 & HUAWEIP8 & ANDROID6 & CHINA & MIXED3 & FULL \\
EXP25 & HTCONEM9 & ANDROID6 & EU & MIXED1 & NORMAL \\
EXP26 & HTCONEM9 & ANDROID6 & EU & MIXED1 & FULL \\
EXP27 & HTCONEM9 & ANDROID6 & EU & MIXED2 & NORMAL \\
EXP28 & HTCONEM9 & ANDROID6 & EU & MIXED2 & FULL \\
EXP29 & HTCONEM9 & ANDROID6 & EU & MIXED3 & NORMAL \\
EXP30 & HTCONEM9 & ANDROID6 & EU & MIXED3 & FULL \\
EXP31 & HTCONEM9 & ANDROID6 & CHINA & MIXED1 & NORMAL \\
EXP32 & HTCONEM9 & ANDROID6 & CHINA & MIXED1 & FULL \\
EXP33 & HTCONEM9 & ANDROID6 & CHINA & MIXED2 & NORMAL \\
EXP34 & HTCONEM9 & ANDROID6 & CHINA & MIXED2 & FULL \\
EXP35 & HTCONEM9 & ANDROID6 & CHINA & MIXED3 & NORMAL \\
EXP36 & HTCONEM9 & ANDROID6 & CHINA & MIXED3 & FULL \\
EXP37 & LGNEXUS & ANDROID6 & EU & MIXED1 & NORMAL \\
EXP38 & LGNEXUS & ANDROID6 & EU & MIXED1 & FULL \\
EXP39 & LGNEXUS & ANDROID6 & EU & MIXED2 & NORMAL \\
EXP40 & LGNEXUS & ANDROID6 & EU & MIXED2 & FULL \\
EXP41 & LGNEXUS & ANDROID6 & EU & MIXED3 & NORMAL \\
EXP42 & LGNEXUS & ANDROID6 & EU & MIXED3 & FULL \\
EXP43 & LGNEXUS & ANDROID6 & CHINA & MIXED1 & NORMAL \\
EXP44 & LGNEXUS & ANDROID6 & CHINA & MIXED1 & FULL \\
EXP45 & LGNEXUS & ANDROID6 & CHINA & MIXED2 & NORMAL \\
EXP46 & LGNEXUS & ANDROID6 & CHINA & MIXED2 & FULL \\
EXP47 & LGNEXUS & ANDROID6 & CHINA & MIXED3 & NORMAL \\
EXP48 & LGNEXUS & ANDROID6 & CHINA & MIXED3 & FULL \\
EXP49 & SAMSUNGS6EDGE & ANDROID6 & EU & MIXED1 & NORMAL \\
EXP50 & SAMSUNGS6EDGE & ANDROID6 & EU & MIXED1 & FULL \\
EXP51 & SAMSUNGS6EDGE & ANDROID6 & EU & MIXED2 & NORMAL \\
EXP52 & SAMSUNGS6EDGE & ANDROID6 & EU & MIXED2 & FULL \\
EXP53 & SAMSUNGS6EDGE & ANDROID6 & EU & MIXED3 & NORMAL \\
EXP54 & SAMSUNGS6EDGE & ANDROID6 & EU & MIXED3 & FULL \\
EXP55 & SAMSUNGS6EDGE & ANDROID6 & CHINA & MIXED1 & NORMAL \\
EXP56 & SAMSUNGS6EDGE & ANDROID6 & CHINA & MIXED1 & FULL \\
EXP57 & SAMSUNGS6EDGE & ANDROID6 & CHINA & MIXED2 & NORMAL \\
EXP58 & SAMSUNGS6EDGE & ANDROID6 & CHINA & MIXED2 & FULL \\
EXP59 & SAMSUNGS6EDGE & ANDROID6 & CHINA & MIXED3 & NORMAL \\
EXP60 & SAMSUNGS6EDGE & ANDROID6 & CHINA & MIXED3 & FULL \\
\midrule
EXP61 & SAMSUNGS6EDGE & ANDROID7 & EU & MIXED1 & NORMAL \\
EXP62 & SAMSUNGS6EDGE & ANDROID7 & EU & MIXED1 & FULL \\
EXP63 & SAMSUNGS6EDGE & ANDROID7 & EU & MIXED2 & NORMAL \\
EXP64 & SAMSUNGS6EDGE & ANDROID7 & EU & MIXED2 & FULL \\
EXP65 & SAMSUNGS6EDGE & ANDROID7 & EU & MIXED3 & NORMAL \\
EXP66 & SAMSUNGS6EDGE & ANDROID7 & EU & MIXED3 & FULL \\
EXP67 & SAMSUNGS6EDGE & ANDROID7 & CHINA & MIXED1 & NORMAL \\
EXP68 & SAMSUNGS6EDGE & ANDROID7 & CHINA & MIXED1 & FULL \\
EXP69 & SAMSUNGS6EDGE & ANDROID7 & CHINA & MIXED2 & NORMAL \\
EXP70 & SAMSUNGS6EDGE & ANDROID7 & CHINA & MIXED2 & FULL \\
EXP71 & SAMSUNGS6EDGE & ANDROID7 & CHINA & MIXED3 & NORMAL \\
EXP72 & SAMSUNGS6EDGE & ANDROID7 & CHINA & MIXED3 & FULL \\
\bottomrule
\end{tabular}
\end{table}

\section{Experimental results}
\label{sec:results}

We analyze software aging phenomena with respect to the research questions of Section~\ref{sec:research_questions}, using the metrics and the experimental plan presented in Section~\ref{sec:methodology}. We conclude the analysis with a more detailed study of software aging symptoms in Section~\ref{subsec:indepth_analysis}. The dataset and results are available at \url{https://doi.org/10.6084/m9.figshare.12136134.v1}.

\subsection{Software aging across Android vendors}
\label{subsec:vendors}

The following experiments aim to assess whether software aging exists in commercial Android devices, and whether aging symptoms vary across different devices. We consider devices from several Android vendors, and we fix the version of the Android OS to 6.

\subsubsection{\textbf{Analysis of Launch Time}}

\begin{figure}[!t]
    \centering
    \includegraphics[width=0.8\textwidth]{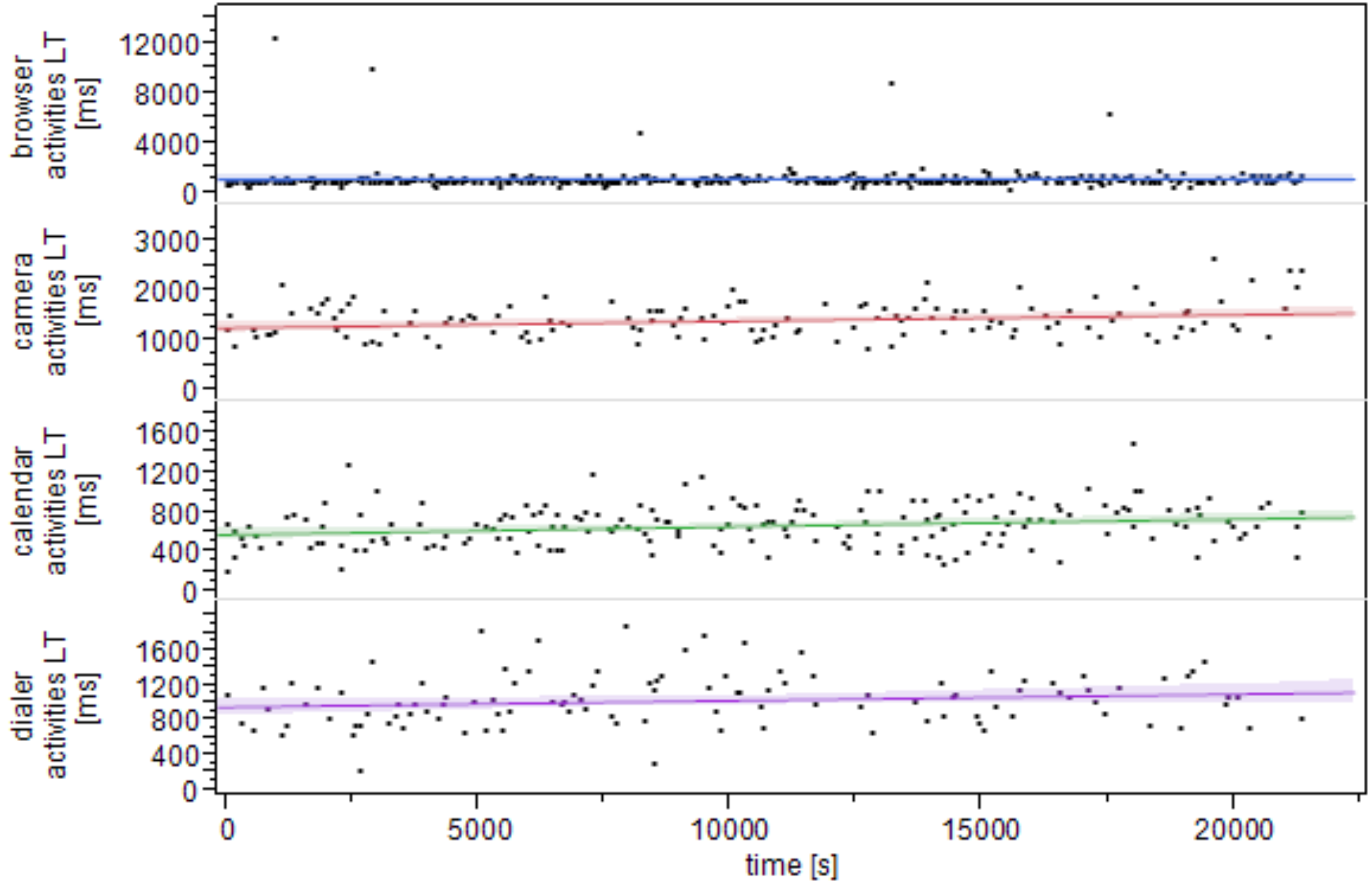}
    \caption{Groups Activities Launch Time for \emph{EXP39}}
    \label{fig:lt_trend_anexperiment}
\end{figure}

To provide insights about the software aging problems found by the experiments, we show in \figurename{}~\ref{fig:lt_trend_anexperiment} examples of LT measurements and trends. The figure focuses on a subset of activities from the experiment \emph{EXP39}, where the activities are divided among ``browser'', ``camera'', ``calendar'', and ``dialer''. 
In the figure, the data points are the LT measurements ; the lines that cross the data points represent the median slope estimated by the Sen's procedure; the colored bands around the lines represent the 95\% confidence interval for the slope according to the Sen's procedure. Since the aging phenomenon takes some hours to manifest, and since there are unavoidable random fluctuations and sporadic outliers in the performance measurements, these trends are not immediately apparent from visual inspection, which is the reason why statistical analysis is needed to address the research questions, such as the Mann-Kendall (MK) trend detection test (see subsection~\ref{subsec:user_perceived}).

\begin{figure*}[!t]
    \centering
    \includegraphics[width=\textwidth]{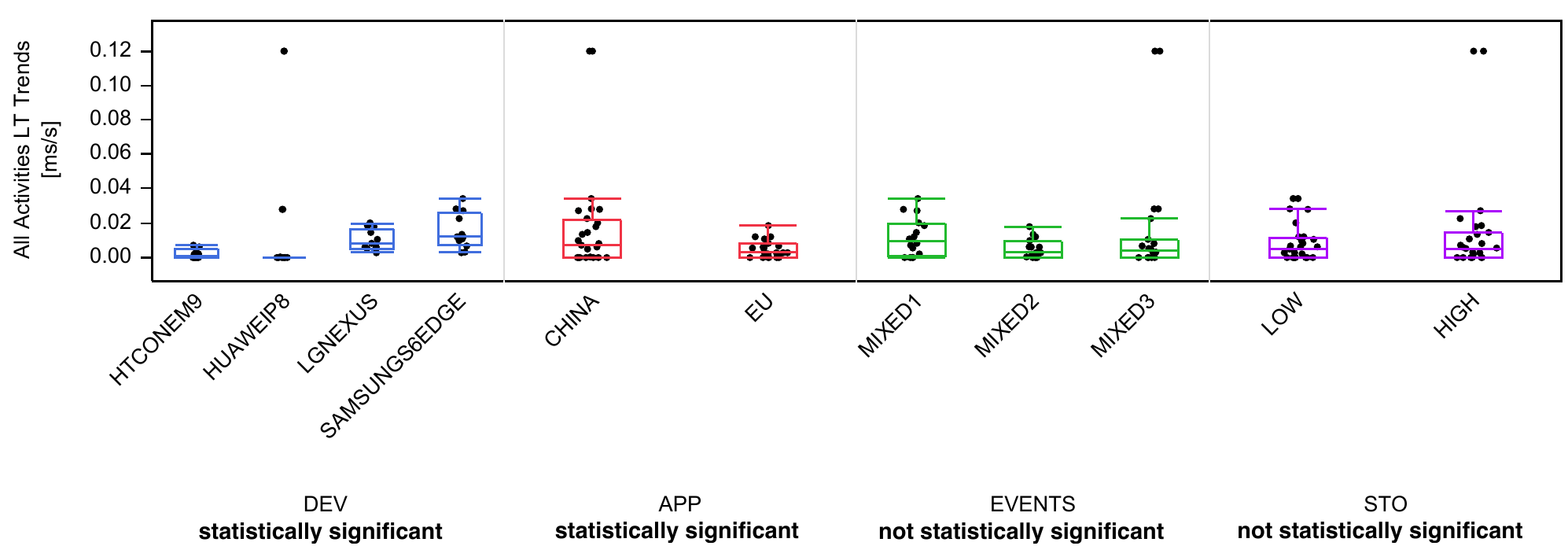}
    \caption{Distribution of the Launch Time Trends, with all vendors and fixed to Android 6 (EXP13$\sim$EXP60)}
    \label{fig:android6_all_activities_lt}
\end{figure*}

We first analyze the  experiments to address our first basic research question, that is, whether software aging phenomena can be found in the Android OS. Indeed, this was the case for all of the tested devices: 
in the majority of the experiments (31 out of 48), the Mann-Kendall test confirmed the presence of a statistically-significant increasing trend in the LT series across all the activities launched during the experiment, which implies that the launch times become longer and longer over time. 
For 29 out of these 31 cases, at least two further tests among the Cox-Stuart, \textit{t-}test and Spearman's rho, confirmed the presence of a trend. 
These trends are summarized in \figurename~\ref{fig:android6_all_activities_lt}. The data points in the figure represent the slopes of the performance degradation trends obtained by the Sen's procedure, in terms of milliseconds of launch time lost per second. The values greater than zero represent the cases were the series exhibited an increasing trend. In total, there are 48 samples, one for each combination of the factors (EXPs 13 to 60 in \tablename{}~\ref{table:experimental_plan}, where the OS version is Android 6). All of the sub-plots show the same data samples from different perspectives, where we split the data samples with respect to different factors (device, application set, workload events, storage space usage).
On average, there was an estimated degradation of 380ms of the launch time after 6 hours of testing (i.e., the additional delay for launching apps after that the device has been stress-tested); in the worst case, there was an estimated degradation of 2.5 seconds of the launch time after 6 hours. At the end of some experiments, the devices were so bloated to be unusable, as they reacted to user inputs with very long delays (e.g., more than 10 seconds).

Moreover, we found that the extent of software aging varies across devices from different vendors. Even if the devices all run Android 6, the vendors introduce their own changes to the open-source version, and these changes can have an impact in terms of software aging. 
We applied the one-way ANOVA technique to assess whether the experimental factors (see \tablename{}~\ref{table:factors_and_levels}) had an influence on the extent of the performance degradation trends. 
\tablename{}~\ref{ANOVAassumptions} reports the detailed results of both the assumptions verification and the p-value for all the ANOVA tests for this work.
According to the ANOVA, the device vendor (\emph{DEV}) is a factor that determines statistically-significant differences in the Launch Time, with a high confidence level (p-value = 0.0460). 
By means of visual analysis of the boxplots, we can determine that the experiments with \emph{HUAWEIP8} yielded the lowest trends, while the \emph{SAMSUNGS6EDGE} yielded the highest ones. More importantly, another statistically-significant difference is in the \emph{APP} factor, with a confidence greater than 99\% (p-value = 0.0044), where the \emph{CHINA} applications have a worse impact than \emph{EU} in terms of LT. This result suggests that software aging in the Android OS depends on the workload, which can stress different services and subsystems of the Android OS depending on user applications. 
Moreover, we found that the extent of software aging varies across different devices, as the vendors adopt different flavors of the Android OS with their own proprietary customizations.



\begin{table*}[]
\centering
\fontsize{6}{7}\selectfont
\caption{Analysis of Variance w.r.t. Normality and Homoscedasticity}
\label{ANOVAassumptions}
\begin{tabular}{cl}
\toprule
$H_{0}$ & all groups have the same effect on the response variable.\\
$H_{alt}$ & the differences in observed effects between groups are unlikely to be due to random chance.\\
$\alpha$ & 0.05 (\ie we reject $H_{0}$ and accept $H_{alt}$ when p-value is lower than 0.05)\\
\midrule
NORM. & normality (YES if the samples came from a normally distributed population, NO otherwise)\\
HOM. & homoscedasticity (YES if the population variances are equal, NO otherwise)\\
\bottomrule\\
\end{tabular}

\begin{tabular}{cccccccc}
\toprule
                                              &         & \multicolumn{2}{c}{SHAPIRO-WILK} & \multicolumn{2}{c}{LEVENE}   &                &                \\
                                              &  \emph{grouped by}   & P\textless W         & NORM.         & P\textgreater F & HOM. & ANOVA & p-value        \\\toprule
\multirow{4}{*}{\makecell{ANDROID 6 \\ LT}}                
					    & DEVICE  & \textless.0001         & NO             & 0.0{078}             & NO       & K-W & {0.0460} \\
                                              & APP     & \textless.0001         & NO             & 0.0{861}            & YES       & K-W & 0.{0044}         \\
                                              & WL      & \textless.0001         & NO             & 0.{3325}            & YES      & K-W & 0.{1428}         \\
                                              & FS      & \textless.0001         & NO             & 0.{3905}            & YES      & K-W & 0.8{744}          \\\midrule
\multirow{4}{*}{\makecell{ANDROID 6 \\ PSS systemserver}}   
					    & DEVICE  & \textless.0001         & NO             & \textless.0001    & NO       & K-W & 0.0{002}         \\
                                              & APP     & \textless.0001         & NO             & {0.0142}    & NO       & K-W & 0.00{33}         \\
                                              & WL      & \textless.0001         & NO             & 0.{9672}            & YES      & K-W & 0.{1953}          \\
                                              & FS      & \textless.0001         & NO             & 0.{8444}            & YES      & K-W & 0.{8510}         \\\midrule
\multirow{4}{*}{\makecell{ANDROID 6 \\ PSS surfaceflinger}} 
					    & DEVICE  & \textless.0001         & NO             & {0.01346}   & NO       & K-W & {0.0003} \\
                                              & APP     & \textless.0001         & NO             & 0.0{363}            & NO       & K-W & 0.0{114}         \\
                                              & WL      & \textless.0001         & NO             & 0.{6004}             & YES      & K-W & 0.{6274}          \\
                                              & FS      & \textless.0001         & NO             & 0.{2157}            & YES       & K-W & 0.{6966}         \\\midrule
\multirow{4}{*}{\makecell{HUAWEI \\ LT}}                    
					    & VERSION & {0.0057}         & NO             & 0.{0562}            & YES      & K-W & 0.{8961}         \\
                                              & APP     & {0.0005}         & NO             & 0.{1679}            & YES      & K-W & 0.{1325}         \\
                                              & WL      & {0.0037}         & NO             & 0.{6114}            & YES      & K-W & 0.{9227}         \\
                                              & FS      & {0.0025}         & NO             & 0.{3738}            & YES      & K-W & 0.{1649}         \\\midrule
\multirow{4}{*}{\makecell{HUAWEI \\ PSS systemserver}}      
					    & VERSION & 0.2{046}                 & YES            & 0.0{110}            & NO       & WELCH          & 0.00{40}         \\
                                              & APP     & 0.{1623}                 & YES            & 0.00{79}            &  NO       & WELCH          & 0.{1714}         \\
                                              & WL      & 0.0{754}                 & YES            & 0.64{65}            & YES      & FISHER         & 0.{5436}         \\
                                              & FS      & 0.0{089}                 & {NO}            & 0.{7904}            & YES      & {K-W}         & 0.{7055}         \\\midrule
\multirow{4}{*}{\makecell{HUAWEI \\ PSS surfaceflinger}}    
					    & VERSION & 0.{1095}                 & {YES}             & 0.{5162}            & YES      & {FISHER} & 0.0{144}          \\
                                              & APP     & 0.{6426}                 & YES            		& 0.{8697}            & YES      & FISHER         & 0.{2736}         \\
                                              & WL      & 0.{1107}                 & {YES}             & 0.8{754}            & YES      & {FISHER} & 0.{8222}         \\
                                              & FS      & 0.{1884}                 & {YES}             & 0.{6679}            & YES      & {FISHER} & 0.{7893}          \\\midrule
\multirow{4}{*}{\makecell{SAMSUNG \\ LT}}                   
					    & VERSION & \textless.0001         & NO             & 0.{1500}            & {YES}       & K-W & 0.{5089}         \\
                                              & APP     & \textless.0001         & NO             & 0.{4255}            & {YES}       & K-W & 0.{3218}         \\
                                              & WL      & \textless.0001         & NO             & 0.{6222}              & {YES}       & K-W & 0.{4657}         \\
                                              & FS      & \textless.0001         & NO             & 0.{4526}            & YES      & K-W & 0.{2207}         \\\midrule
\multirow{4}{*}{\makecell{SAMSUNG \\ PSS systemserver}}     
					    & VERSION & 0.{1368}                  & YES            & {0.0016}    & NO       & WELCH          & 0.0{220}         \\
                                              & APP     & 0.2{010}                 & YES            & \textless.0001    & NO       & WELCH          & 0.00{23}        \\
                                              & WL      & 0.000{3}                 & NO             & 0.{9196}             & YES      & K-W & 0.{8692}         \\
                                              & FS      & 0.000{5}                 & NO             & 0.{5976}            & YES      & K-W & 0.{9439}         \\\midrule
\multirow{4}{*}{\makecell{SAMSUNG \\ PSS surfaceflinger}}   
					    & VERSION & 0.0{224}                 & NO             & 0.00{39}            & NO       & K-W & 0.{2353}         \\
                                              & APP     & 0.19{38}                 & YES            & {0.0634}    & {YES}       & {FISHER}          & 0.0{330}         \\
                                              & WL      & 0.00{18}                 & NO             & 0.{9644}            & YES      & K-W & 0.2{437}         \\
                                              & FS      & 0.00{27}                 & NO             & 0.{5378}            & YES      & K-W & 0.{8696}        \\\bottomrule
\end{tabular}
\end{table*}

\subsubsection{\textbf{Analysis of Memory Usage}}

Based on the results of our preliminary work \citep{cotroneo2016software}, we focus the analysis on the \emph{PSS} metric collected for four key processes of the Android OS, namely the \emph{System Server}, \emph{Media Server}, \emph{System UI}, and \emph{Surface Flinger}. 
These processes play an important role in the Android OS: 

\begin{itemize}

\item The \emph{System Server} is the first Java process that starts at Android OS boot and initializes the rest of the Android Framework. It runs the majority of system services, such as the \emph{Activity Manager}, which manages the life cycle of applications and their activities, and the \emph{Package Manager}, which manages installed packages and security permissions.

\item The \emph{Media Server} is the process that hosts most of the media related services, \emph{e.g.} \emph{Audio Flinger}, \emph{Media Player Service}, \emph{Camera Service}, and \emph{Audio Policy Service}.

\item The \emph{System UI} is the process that composes notifications, device status, and device navigation buttons as system bar elements in specific screen areas. 

\item The \emph{Surface Flinger} process receives window layers (surfaces) from multiple sources (\emph{System UI} included), combines them, and displays them on the screen.

\end{itemize}

\begin{figure*}[!t]
    \centering
    \includegraphics[width=\textwidth]{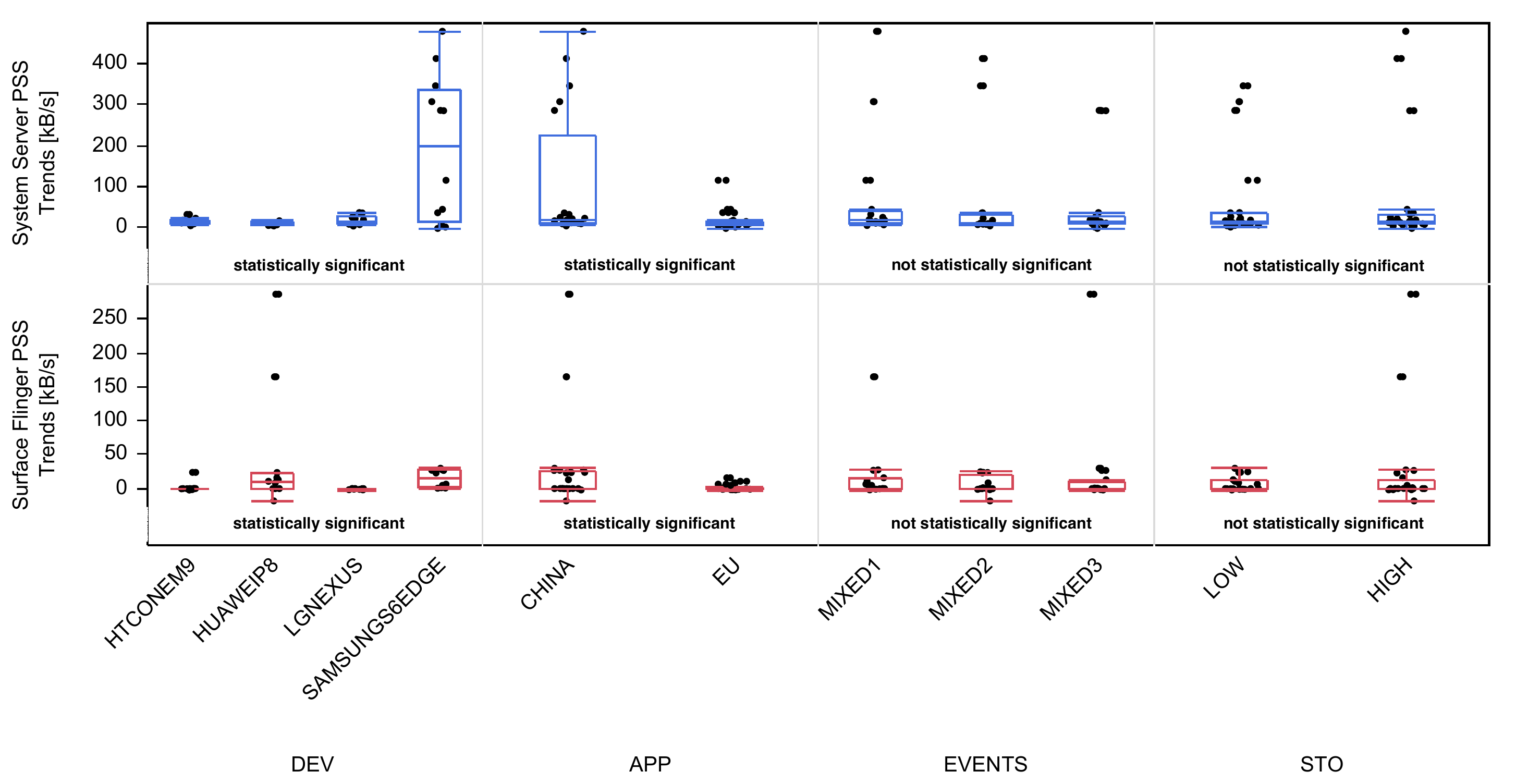}
    \caption{PSS Trends Distributions: EXP13$\sim$EXP60 (Android 6)}
    \label{fig:android6_pss}
\end{figure*}

We again performed the one-way ANOVA, using the PSS of these processes. \figurename{}~\ref{fig:android6_pss} shows the distribution of PSS trends for the \emph{System Server} and the \emph{Surface Flinger}. These two processes exhibited increasing memory consumption trends. In particular, the \emph{System Server} is the process with the highest trends. Moreover, the \emph{DEV} and the \emph{APP} factors have a statistically-significant influence on the \emph{System Server}, with very high confidence (p-values of 0.0002 and 0.0033, respectively): these trends are especially high in the case of the \emph{SAMSUNGS6EDGE}, and of the group of \emph{CHINA} apps. The \emph{EVENTS} and \emph{STO} factors do not have a statistically-significant influence.

The remaining processes (\emph{Media Server} and \emph{System UI}) exhibit negative trends (meaning that, in some cases, memory consumption has been decreasing). Therefore, the memory consumption of these processes does not seem to be related to the performance degradation that we found in the previous analysis. We cross-check this interpretation of the results by jointly analyzing the \emph{PSS} and the \emph{LT} metrics. We compared the memory consumption trend of the processes with the corresponding LT trends of the experiments, by correlating these two metrics using the \emph{Spearman's rank correlation}.

\tablename{}~\ref{table:perprocessmemory} shows the results of the correlation. Indeed, the memory consumption of the \emph{System Server} and the \emph{Surface Flinger} exhibit a noticeable (and statistically significant) positive correlation with the LT, meaning that high LT trends (i.e., quicker performance degradation) occur at the same time of high PSS trends (i.e., quicker inflation of the memory consumption). Instead, the other processes exhibit a less significant correlation, and even negative. Thus, the memory consumption of the \emph{Media Server} and \emph{System UI} does not seem a possible cause of the performance degradation (i.e., the LT trends). Instead, the increasing memory consumption of the \emph{System Server} (which has an important role in starting and managing activities through the \emph{Activity Manager} and \emph{Package Manager}) is a potential symptom of software aging, that we further investigate later in this section (\S{}~\ref{subsec:indepth_analysis}).

\begin{table}[!t]
\caption{Spearman correlation coefficients between All Activities LT trends and PSS trends of Android system processes.}
\label{table:perprocessmemory}
\centering
\begin{tabular}{C{3cm}C{3cm}C{1.8cm}@{}m{0pt}@{}}
\toprule
PROCESS & SPEARMAN COEFFICIENT & P-VALUE&\\
\midrule
system\newline (System Server) & 0.6467 & 0.0001 &\\
\midrule
mediaserver\newline (Media Server) & -0.5054 & 0.0052 & \\
\midrule
com.android.systemui\newline (System UI) & -0.2135 & 0.91 & \\
\midrule
surfaceflinger\newline (Surface Flinger) & 0.6409 & 0.0002 &\\
\bottomrule
\end{tabular}
\end{table}

\subsection{Software aging across Android versions}
\label{subsec:versions}

We analyze software aging across different versions of the Android OS, by looking for differences both between the Android versions 5 and 6 (by locking the \emph{DEV} factor to \emph{HUAWEIP8}), and between the Android versions 6 and 7 (by locking the \emph{DEV} factor to \emph{SAMSUNGS6EDGE}).

\subsubsection{\textbf{Analysis of Launch Time}}

\begin{figure*}[!t]
    \centering
    \includegraphics[width=\textwidth]{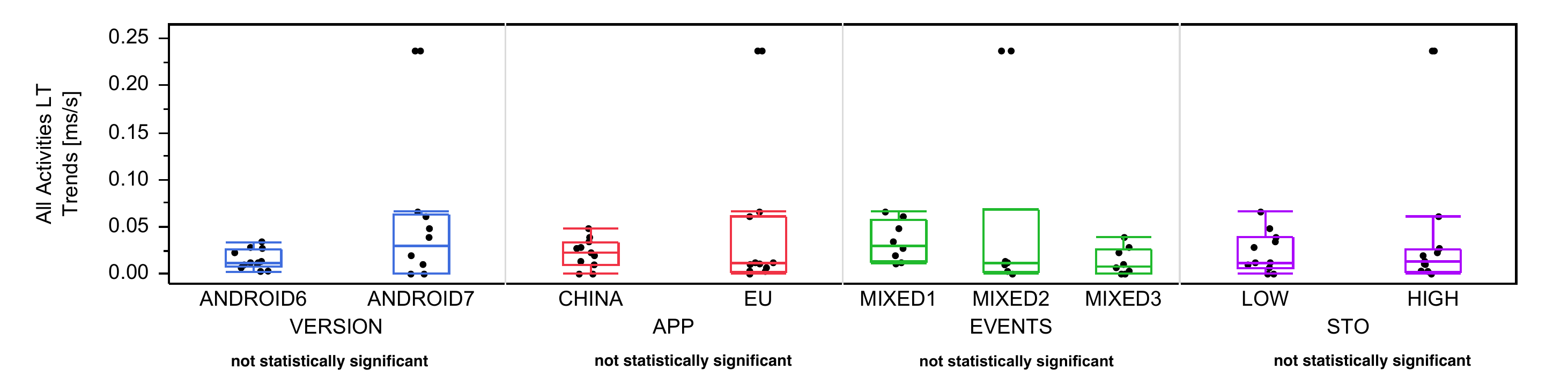}
    \caption{Launch Time Trends Distributions: EXP49$\sim$EXP72 (Samsung S6 Edge)}
    \label{fig:samsungs6edge_all_activities_lt}
\end{figure*}

\begin{figure*}[!t]
    \centering
    \includegraphics[width=\textwidth]{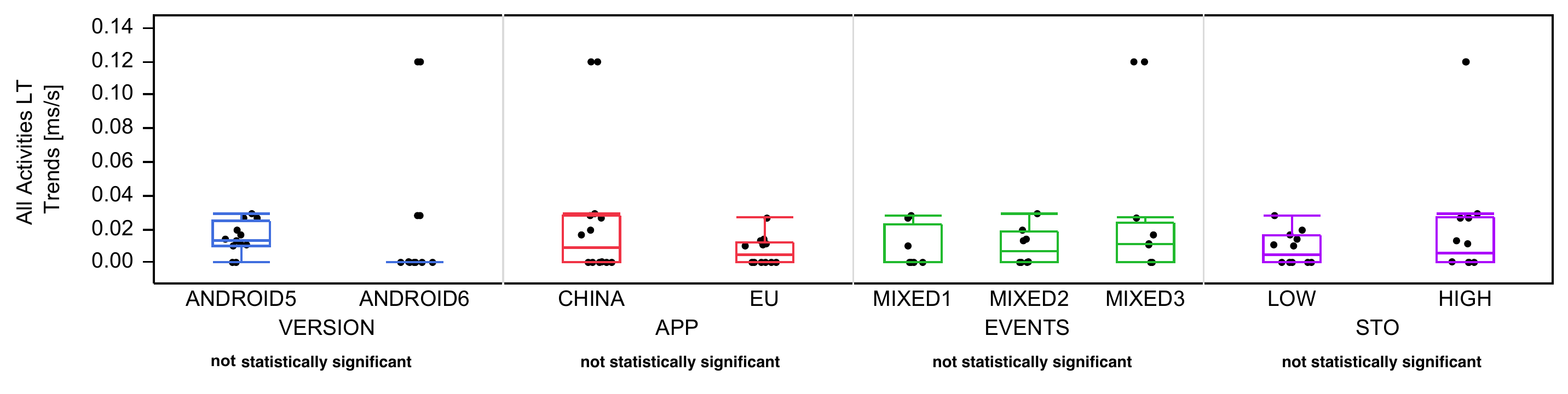}
    \caption{Launch Time Trends Distributions: EXP1$\sim$EXP24 (Huawei P8)}
    \label{fig:huaweip8_all_activities_lt}
\end{figure*}

In the case of the \emph{SAMSUNGS6EDGE} device, we again observed aging trends also for \emph{ANDROID7}, which are shown in \figurename{}~\ref{fig:samsungs6edge_all_activities_lt}. On average, there was an estimated degradation of 650.89ms of the launch time after 6 hours of testing; in the worst case, this estimated performance degradation was up to 5.1 seconds. 
In the case of the \emph{HUAWEIP8} device, we also notice aging trends in both the versions, as shown in \figurename{}~\ref{fig:huaweip8_all_activities_lt}. In this case, the estimated performance degradation at the end of testing has been 291.62ms on average, and 2.6 seconds in the worst case.

Comparing \emph{ANDROID6} with \emph{ANDROID7} (\figurename{}~\ref{fig:samsungs6edge_all_activities_lt}), the LT trends show only small differences in the mean values, and slightly higher variability of the trends for \emph{ANDROID7}. The differences between \emph{ANDROID5} and \emph{ANDROID6} (\figurename{}~\ref{fig:huaweip8_all_activities_lt}) are apparently more noticeable, with a reduction of the LT trends in favor of \emph{ANDROID6}. We performed the one-way ANOVA on these two experiments, to assess whether the Android version has a statistically-significant influence. 
According to the ANOVA, the Android version had no influence in both \emph{HUAWEIP8} and \emph{SAMSUNGS6EDGE}. This result suggests that the revisions to the Android OS are not addressing the areas of the OS that are affected by software aging, and that Android vendors need to invest more effort to address this neglected problem.

\subsubsection{\textbf{Analysis of Memory Usage}}

\begin{figure}[!t]
    \centering
    \begin{subfigure}[b]{0.44\textwidth}
        \includegraphics[width=\textwidth]{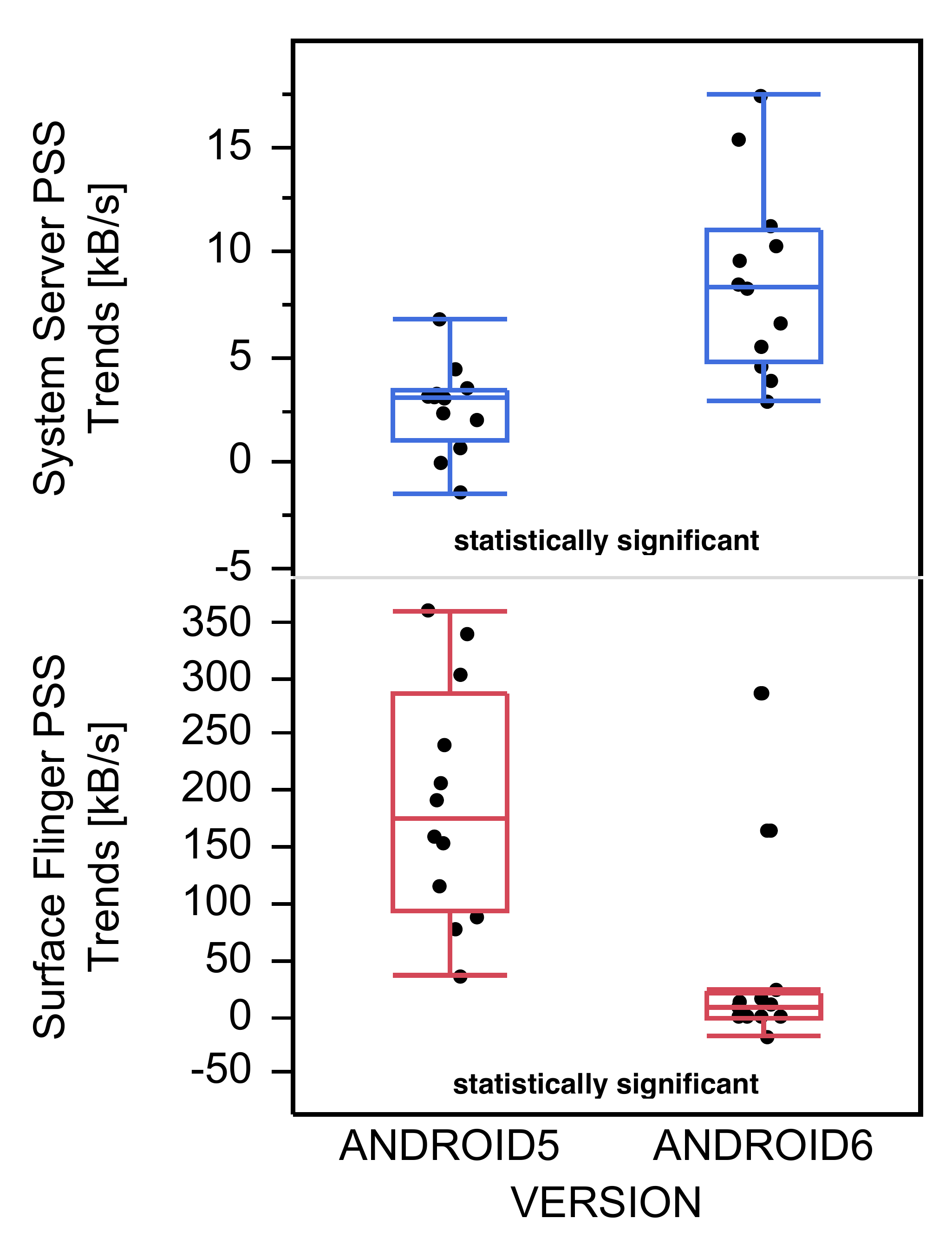}
        \caption{EXP1$\sim$EXP24 (Huawei P8)}
        \label{fig:huawei_pss}
    \end{subfigure}
    \hfill
    \begin{subfigure}[b]{0.44\textwidth}
        \includegraphics[width=\textwidth]{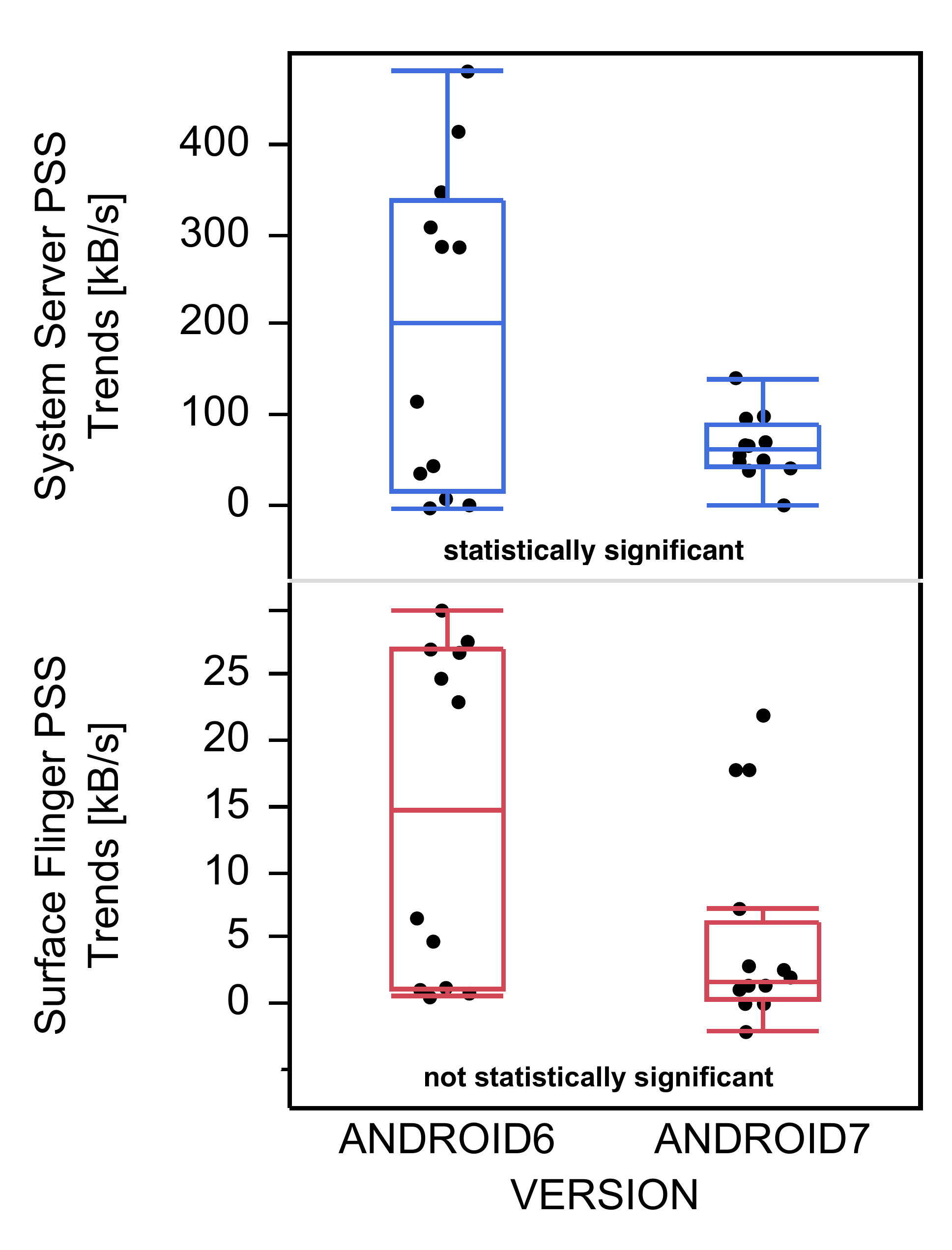}
        \caption{EXP49$\sim$EXP72 (Samsung S6 Edge)}
        \label{fig:samsung_pss}
    \end{subfigure}
    \caption{PSS Trends Distributions: }\label{fig:pss_trends}
\end{figure}

Figure~\ref{fig:huawei_pss} and Figure~\ref{fig:samsung_pss} show the PSS trends for the \emph{System Server} and \emph{Surface Flinger} processes, respectively in the case of \emph{HUAWEIP8} and \emph{SAMSUNGS6EDGE}. 
According to the ANOVA, there were statistically-significant differences between the trends of different Android OS versions. 
In particular, in the case of \emph{HUAWEIP8} (i.e., the transition from \emph{ANDROID5} to \emph{ANDROID6}), the PSS trends for the \emph{System Server} process gets worse; instead, in the case of the \emph{SAMSUNGS6EDGE} (i.e., the transition from \emph{ANDROID6} to \emph{ANDROID7}), the PSS trends for the \emph{System Server} exhibit an improvement. 
Considering the results of the previous analysis on LT trends, it seems that the magnitude of LT trends is not impacted by these variations of the PSS trends (i.e., the LT trends are steady even if the PSS trends are different). 
This result suggests that, probably, it is not only the quantity of memory consumption that influences the performance degradation, but also the way the memory is managed, in terms of fragmentation and lifespan of the objects (see also \S{}~\ref{subsec:garbage_collection}). We analyze this aspect in more detail in the next sections.

\subsection{Analysis of process internals}
\label{subsec:indepth_analysis}

In this section, we analyze the Garbage Collection, and task-related events, which provide more information about the internal behavior of Android processes, to get more insights about the reasons of the aging trends discussed in the previous sections.

\subsubsection{\textbf{Analysis of Garbage Collection}}
\label{subsec:gc}

As discussed in subsection~\ref{subsec:garbage_collection}, we check how many experiments exhibited an anomalous garbage collection activity, as it is an indicator that the software aging problems are related to memory management problems, such as memory bloat (e.g., stale objects that are still referenced and are not collectable) and memory fragmentation.

As metrics for GC activity, we consider the \emph{GC Pause Time} (i.e., the period that the process is suspended during the GC), and the \emph{GC Duration} (i.e., the total duration, including both the GC that executes when the process is suspended, and the GC that executes in parallel with the program). We separately analyze the different types of GC that are performed by the Android Run-Time \citep{android-web-debugging-memory}: in particular, in our experiments, only two types of GC produced more than 100 samples and exhibited some trend with confidence higher than 95\%, namely \emph{Concurrent} GC (which is performed by a separate thread in background) and \emph{Explicit} GC (where GC is performed on explicit request of a process).

\begin{figure*}[!th]
    \centering
    \includegraphics[width=\textwidth]{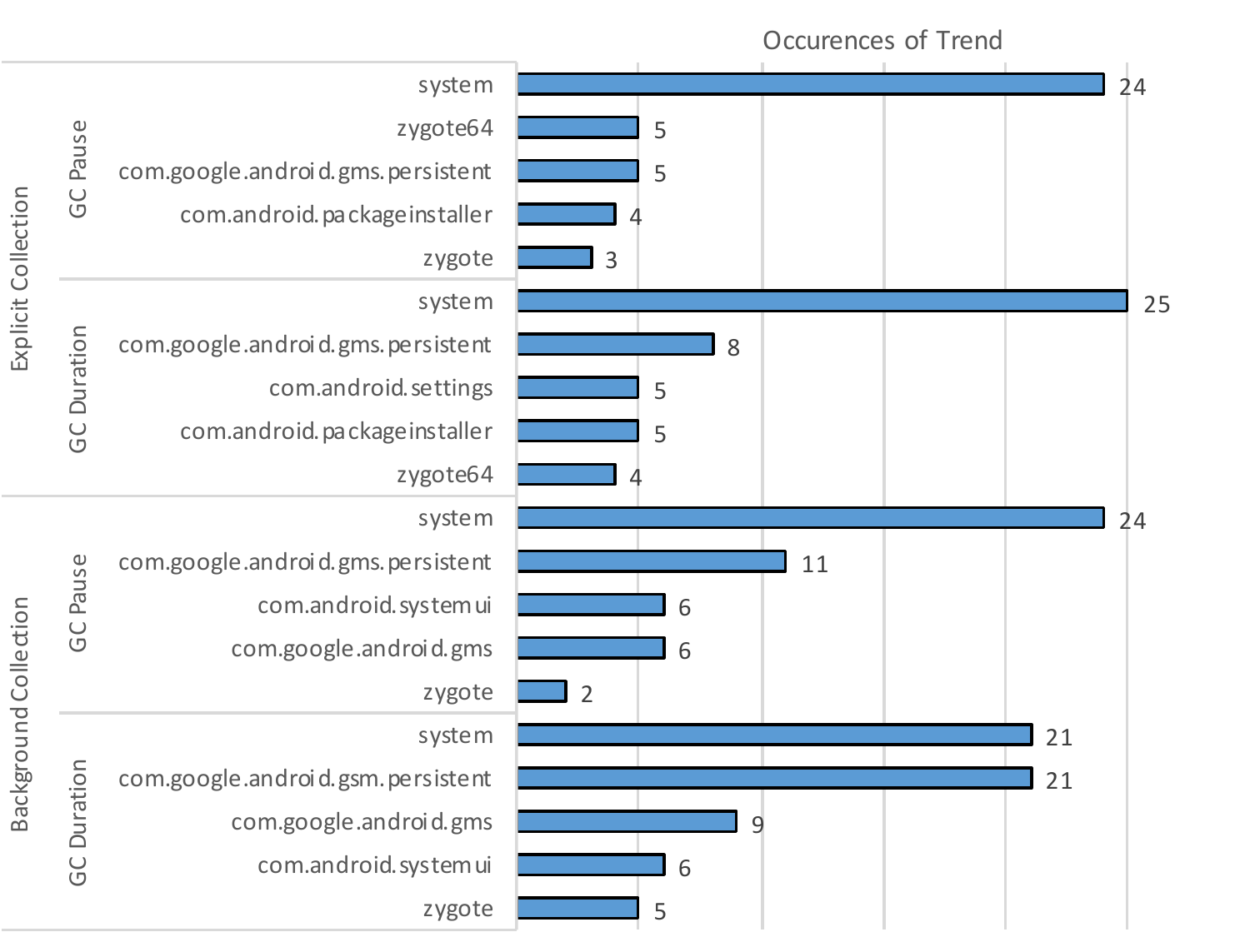}
    \caption{Top five processes in EXP1$\sim$EXP72 that exhibit the greatest number of trend occurrences in each of the 4 garbage collection metrics (GC Duration, GC Pause Time, Explicit Collection, and Background collection times).}
    \label{fig:gc_rank}
\end{figure*}

We separately analyze each process of the Android OS, and each of the four GC metrics (GC Duration/Pause Time, and Explicit/Background Collection). We counted the number of experiments in which the garbage collection exhibited a statistically-significant increasing trend. These counts are showed in \figurename~\ref{fig:gc_rank}.
In the figure, we ranked the processes with respect to these values, and showed the 5 processes that most frequently exhibited a trend in the GC pause times and GC durations. From the experiments, we noted that such trends occurred most often in the System Server process (which is internally labeled as \emph{system} in the Android OS). This result suggests that this process heavily uses heap memory, and that it is exposed to performance degradation due to the inflation and fragmentation of the heap, which increases the overhead of garbage collections and slows down or freezes its threads. 
The Android OS is sensitive to slow-downs of this process: the System Server provides key services for managing the applications' lifecycle (such as the Activity Manager and the Package Manager, which both run as threads inside the System Server, and which are invoked every time that an application is started, disposed of, etc.). Since garbage collection delays the System Server (by freezing or slowing down the process for a short amount of time), these delays propagate to other Android components and affect the responsiveness of the device. 
Moreover, this result suggests that monitoring the GC times of these processes is another useful indicator to detect software aging in the Android OS, and can be leveraged to trigger software rejuvenation actions when GC becomes too slow or too frequent.

\subsubsection{\textbf{Analysis of task-level metrics}}
\label{subsec:task}

Following up on our previous analysis of Android OS processes, we perform a finer-grain analysis of software aging at \emph{task} level (see also \ref{subsec:task_level_metrics}). This analysis is useful since the processes of the Android OS contain up to hundreds of tasks to manage multiple services. For example, both the Activity Manager and the Package Manager run within the System Server process, together with tens of other services.  
Since process-level metrics (such as the PSS) cannot provide information about individual tasks, we analyze task-level metrics for specific Android services.

For each experiment, and for each group of tasks, we apply the MK trend detection test on the time series of \emph{major faults}, \emph{minor faults}, \emph{kernel time}, and \emph{user time}, as defined in subsection~\ref{subsec:task_level_metrics}. Then, we counted the number of experiments in which the task group (\ie at least one task in the group) exhibited an increasing utilization of CPU and memory, and ranked the groups according to the count.

\begin{figure}[!th]
    \centering
    \begin{subfigure}[b]{0.95\textwidth}
        \includegraphics[width=\textwidth]{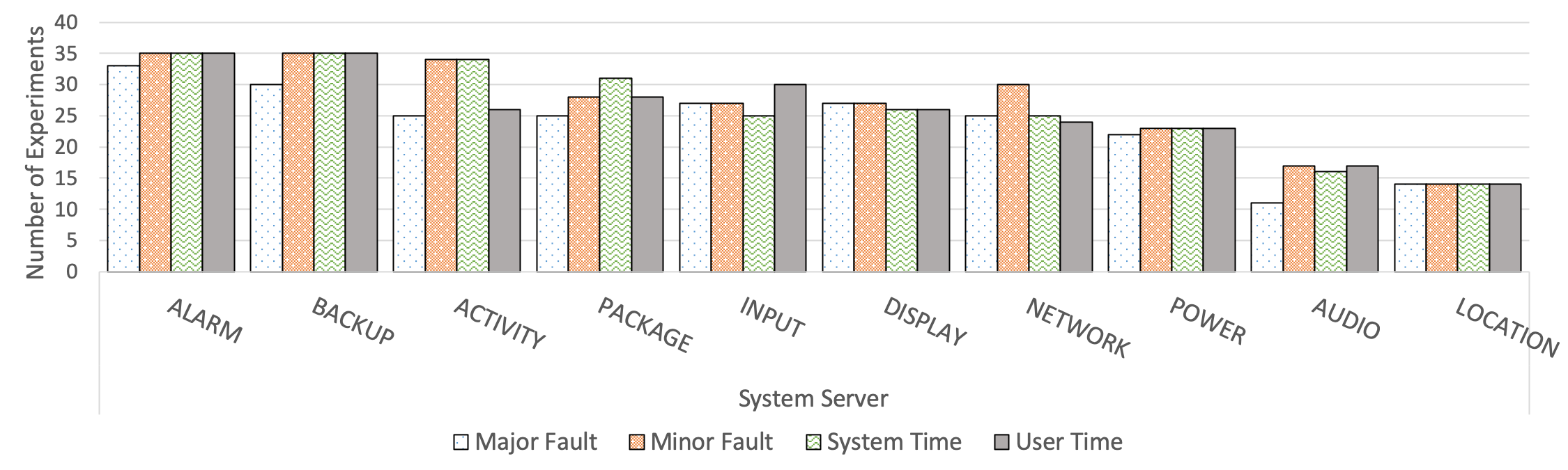}
        \caption{System Server}
        \label{fig:tasks_rank_system_server}
    \end{subfigure}
    \begin{subfigure}[b]{0.95\textwidth}
        \includegraphics[width=\textwidth]{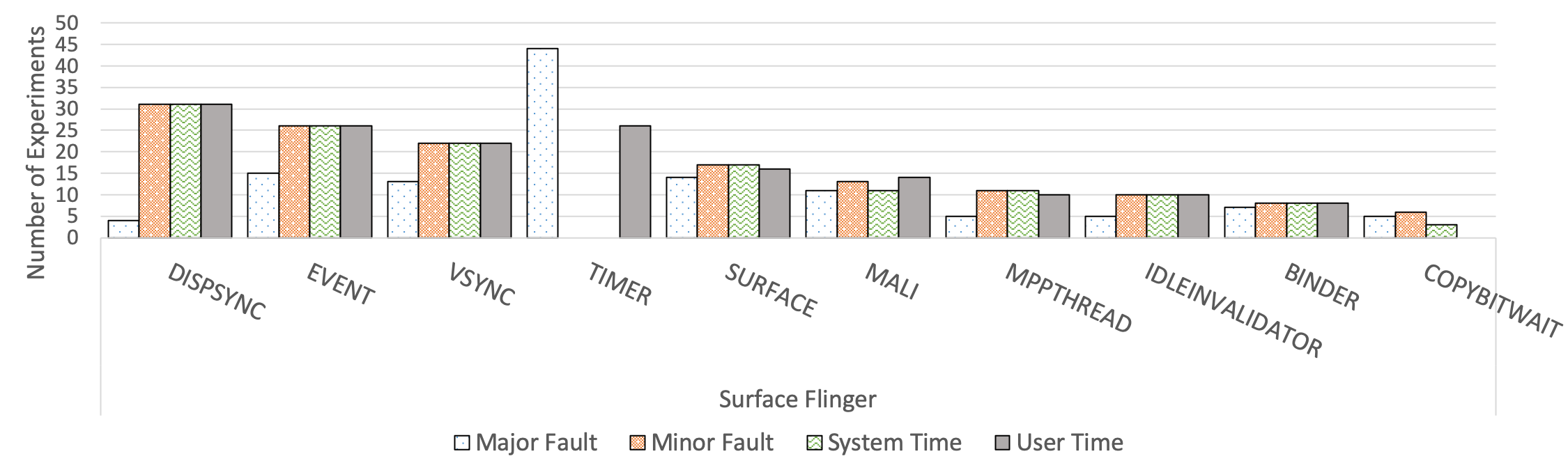}
        \caption{Surface Flinger}
        \label{fig:tasks_rank_surface_flinger}
    \end{subfigure}
    \caption{Top 10 task groups that exhibit the highest number of task-level metric trends in EXP1$\sim$EXP72, respectively in the System Server (a) and Surface Flinger (b) processes. Major and minor faults are memory accesses served respectively from the storage, and from the page cache or other processes. User Time and System Time are the CPU time spent by a task running application- (e.g., for CPU-bound activity) or kernel-code (e.g., for I/O activity).}
    \label{fig:tasks_rank}
\end{figure}

According to the previous results, we focus on task-level metrics for the \emph{System Server} and the \emph{Surface Flinger}.
To relate the tasks to Android subsystems, we grouped the tasks with respect to the Android service or subsystem they belong to, according to their names and to our analysis of the Android AOSP source code. For example, the \emph{ACTIVITY} group in \emph{System Server} consists of four threads related to the Activity Manager, namely \emph{ActivityManager}, \emph{ActivityManager\_2}, \emph{ActivityManger\_3}, and \emph{HwActivityManag}. The value of each group is computed by averaging the trends count of each task of the group. \figurename~\ref{fig:tasks_rank} shows the 10 groups with the highest ranking.

In the case of the \emph{System Server}, the task groups that most frequently exhibit trends include:

\begin{itemize}

\item \emph{ALARM}: the tasks that execute the \emph{Alarm Manager} service, that is in charge of setting up timers for the rest of the system.
\item \emph{BACKUP}: the tasks that execute the \emph{Backup Manager}, which is notified each time there is new data to be saved persistently (\eg new contacts in the dialer).
\item \emph{ACTIVITY}: the tasks that execute the \emph{Activity Manager} service, which handles requests for managing the lifecycle of Android activities.
\item \emph{PACKAGE}: the tasks that execute the \emph{Package Manager} service, which handles requests for forwarding intents and checking permissions.
\item \emph{INPUT}: the tasks that read and dispatch user inputs from the hardware devices to the higher layers.
\item \emph{NETWORK}: the tasks for scanning and for handling the connection to WiFi networks.

\end{itemize}

These groups and the others in \figurename{}~\ref{fig:tasks_rank} represent Android OS tasks that consumed increasing amounts of CPU time (e.g., due to the increasing amount of time spent on traversing bloated data structures, or on garbage collection). For example, the ``Activity'' and ``Alarm'' groups show an increasing trend of task-level metrics, which we attribute to memory bloat that develops in these services due to the very large number of activities and notifications generated during the experiments. These tasks are suspected to be causing software aging, and we further analyze them in the next subsection. 
Moreover, this information points out areas of the Android OS that may be targeted by software rejuvenation: in particular, it is advisable to focus software rejuvenation in the \emph{System Server}, either at process-level (in order to rejuvenate all of the tasks inside the process) or at task-level (by re-initializing the top-most services in the ranks, in order not to disrupt other services inside the process).

\subsection{An experiment with software rejuvenation to counteract memory bloat}
\label{subsec:rejuvenation}

According to the previous analyses, the performance degradation of the Android devices was accompanied by higher memory consumption of key Android OS processes (in particular, the \emph{system server} process), and the increasing of garbage collection activity (i.e., longer duration of garbage collections, and longer suspensions of Android OS processes). Moreover, the analysis pointed out that specific Android OS tasks were consuming more and more CPU time. From these experimental facts, we hypothesize that software aging has been mostly caused by bloated Java containers, which is a recurring issue in large Java projects \citep{xu2008precise,xu2010detecting,Ghanavati2019}. 
We performed an additional experiment to verify this hypothesis, in order to gain further insights into the root causes of software aging. In this experiment, we first instrumented the open-source version of the Android OS with a mechanism to periodically rejuvenate bloated Java containers; then, we stressed again an Android device, in order to evaluate whether the lower memory bloat has an impact on software aging of the device.

The software rejuvenation mechanism mitigates memory bloat by forcefully removing the contents of selected Java containers, such as by invoking the \texttt{.clear()} method of the \texttt{ArrayList} class.
We focus rejuvenation on the \emph{system server} process, as the previous analysis showed that this process was the one most affected by aging symptoms, and on its tasks that exhibited anomalous CPU and memory activity. 
For each of these tasks, we inspected the source code of the Android services that are executed by the tasks. For example, in the case of the \emph{Activity Manager} service (which is executed by the tasks labeled as \emph{ACTIVITY} in the previous analysis), we inspected the source code of the \texttt{ActivityManagerService} Java class, as this class provides concrete implementations of APIs of the \emph{Activity Manager} service. These classes are instantiated by a single, long-lived object since the initialization of the \emph{system service}, during the boot of the device. 
In our inspection, we looked for Java containers among member variables of these classes and allocated at object initialization, as these containers are also long-lived and can be potentially affected by memory bloat. Instead, we do not consider short-lived containers (e.g., only limited to the scope of an individual method).

\begin{table}[]
\caption{Java containers selected for rejuvenation}
\label{table:rejuvenated_containers}
\centering
\begin{tabular}{cccccc}
\toprule
\textbf{Name} & \textbf{Type} & \textbf{Service} \\
\midrule
mLog & LinkedList<String> & WifiService \\    
mLogRecVector & Vector<LogRec> & WifiService \\   
mPendingProcessChanges & ArrayList<ProcessChangeItem> & ActivityManager \\
mAvailProcessChanges & ArrayList<ProcessChangeItem> & ActivityManager \\
mPendingUidChanges & ArrayList<UidRecord.ChangeItem>  & ActivityManager \\
mAvailUidChanges & ArrayList<UidRecord.ChangeItem>  & ActivityManager \\
\bottomrule
\end{tabular}
\end{table}

In order to avoid that flushing the container leads to failure of the \emph{system server} process, we also analyze Java containers with disposable data, such as message logs, resource utilization statistics, and information on recently-executed apps. Therefore, the Java containers we selected for rejuvenation (listed in \tablename{}~\ref{table:rejuvenated_containers}) are a conservative subset of the containers used by the service. For example, in the case of the \emph{WifiService}, we identified three \texttt{ArrayList} containers where the service accumulates message logs about scans WiFi networks and other protocol events. Since these data may cause memory fragmentation and are only useful for debugging purposes, we remove them from the container in order to mitigate software aging.

We remark that we do not aim to develop a fully-fledged software rejuvenation solution, but only to perform a controlled experiment: therefore, we focus rejuvenation on containers that could be rejuvenated with very low risk of side effects on the usability of the device. If the device exhibits less software aging when rejuvenating this subset of containers (i.e., lower performance degradation trends), then the experiment provides additional evidence that the software aging effects are caused by bloated Java containers. Moreover, our measured performance improvement is a lower bound of the potential improvement that could be achieved by actual Android OS developers, which have the expertise to identify more opportunities for safe rejuvenation of containers with more selective rejuvenation (e.g., only removing stale objects related to apps that were terminated since a long time, or objects that could be easily recreated if necessary in a later moment).


We implemented and evaluated this software rejuvenation mechanism on the open-source version of Android OS (AOSP), as we needed access to source code to introduce the changes. We ran the experiment on the LG Nexus device, which was the only device that could execute the AOSP among the devices in our testbed, since the other devices require proprietary customizations to the Android OS (e.g., device drivers). We ran the experiment on the AOSP version 6.0.1, using the Chinese set of Android apps, with a uniform distribution of application switches, touch, motions, trackballs, and navigation events, and with default storage space usage. The experiment lasted 7 hours, and rejuvenation was triggered periodically every hour.

As in the previous experiments, we periodically restarted the apps to take measurements of the launch time. To evaluate the impact of rejuvenation on performance, we compare the performance degradation trend slopes respectively with rejuvenation ($LT^r$) and without it ($LT$):

\begin{equation}
\textrm{Gain}_{LT}\% = \frac{LT - LT^r}{LT} \cdot{} 100
\end{equation}

For instance, a gain of 40\% means that the Launch Time after 6 hours is 40\% lower in the with-rejuvenation case compared to the without-rejuvenation case. 
Moreover, we evaluate software rejuvenation with respect to the \emph{Time to Aging Failure} (TTAF), that is, the period of time after which the launch time of an activity exceeds a limit, over which performance is considered degraded; we set the limit at +200ms with respect to the initial launch time of the activity, as 200ms is the performance goal of basic apps of the Android OS \citep[ch. 10.8]{tanenbaum2014modern}, \citep{android-web-perf-anr}. We compute the expected TTAF with and without rejuvenation ($TTAF^r$, $TTAF$) based on the estimated slopes ($LT^r$, $LT$), and evaluate the gain with the following metric: 

\begin{equation}
\textrm{Gain}_{TTAF}\% = \frac{TTAF^r - TTAF}{TTAF} \cdot{} 100
\end{equation}

For instance, if the TTAF without rejuvenation for a given activity was 6 hours, and the TTAF with rejuvenation is prolonged to 8 hours, then the gain is $2/6 \cdot 100 = 33.33 \%$.

The experiment was repeated three times. \tablename{}~\ref{table:rejuvenation_results} reports the averaged results without and with rejuvenation, and the gain metrics in terms of launch time increase and TTAF. Across all activities that exhibited aging trends, the average gain was 46\% in terms of launch time increase, and 87\% in terms of TTAF. These results point out that flushing the Java containers relieves heap memory management and improves the responsiveness of services running in the \emph{system server} process. Thus, the rejuvenated Java containers were indeed affected by memory bloat. Moreover, our rejuvenation mechanism did not cause any side effect, without any high-severity error logged either by the \emph{system server} or by the apps. Thus, this experiment points out that software rejuvenation is a viable strategy, which brings a measurable benefit by postponing the effects of software aging.

{
\begin{sidewaystable}
\caption{Performance degradation trends with and without software rejuvenation in Java containers.}
\label{table:rejuvenation_results}
\centering
\begin{tabular}{p{3.5cm} @{\hskip 0.5in} ccc @{\hskip 0.5in} ccc @{\hskip 0.5in} cc}
\toprule
    &  \multicolumn{3}{l}{\textbf{Without rejuvenation}} & \multicolumn{3}{l}{ ~~ \textbf{With rejuvenation}} & & \\
\emph{Activity}   &    \emph{Slope}   &   \emph{LT increase}   &   \emph{TTAF}    &  \emph{Slope}    &   \emph{LT increase}    &   \emph{TTAF}  &   \emph{Gain$_{LT}$}    &   \emph{Gain$_{TTAF}$}  \\
    &   [ms/s]   &  [ms]   &    [h]   &   [ms/s]   &  [ms]   &    [h]  &   [\%]   &   [\%]  \\
\midrule
com.android \newline .packageinstaller\_ \newline .permission.ui \newline .GrantPermissionsActivity   &   0.002   &   53.660   &   22.363   &   0.000   &   9.582   &   84.435   &   +82\%   &   +278\%  \\
\midrule
com.baidu.searchbox\_ .MainActivity   &   0.008   &   167.181   &   7.178   &   0.001   &   29.303   &   9.507   &   +82\%   &   +32\%  \\
\midrule
com.moji.mjweather\_ \newline .activity.main \newline .AddCityFirstRunActivity   &   0.009   &   197.860   &   6.065   &   0.006   &   134.905   &   9.140   &   +32\%   &   +51\%  \\
\midrule
com.sina.weibo\_ .SplashActivity   &   0.004   &   95.191   &   12.606   &   0.002   &   46.939   &   27.315   &   +51\%   &   +117\%  \\
\midrule
com.UCMobile.intl\_com .uc.browser.InnerUCMobile   &   0.003   &   66.033   &   18.173   &   0.002   &   48.503   &   29.195   &   +27\%   &   +61\%  \\
\midrule
com.youku.phone\_ \newline .ActivityWelcome   &   0.011   &   237.575   &   5.051   &   0.010   &   208.674   &   5.843   &   +12\%   &   +16\%  \\
\midrule
com.youku.phone\_com .youku.ui.activity. \newline HomePageActivity   &   0.010   &   225.135   &   5.330   &   0.007   &   147.786   &   8.425   &   +34\%   &   +58\%  \\
\midrule
\textbf{Average}   &      &      &      &      &      &      &   \textbf{+46\%}   &   \textbf{+87\%}  \\
\bottomrule
\end{tabular}
\end{sidewaystable}
}

\section{Key findings}
\label{sec:conclusion}

We analyzed software aging issues in the Android OS, by performing a large experimental study across several devices, versions, and test configurations. We obtained a large amount of data that provided us information to address the research questions of this work.

\vspace{5pt}
\noindent
\textbf{(RQ1) Is the Android OS affected by software aging effects? } 
We observed software aging effects in the vast majority of our tests, which thus confirms that the Android OS is indeed affected by software aging effects. We found that aging impacted the responsiveness of the device, as confirmed by the increasing, statistically-significant trends of the launch time of Android activities (\cfr \figurename~\ref{fig:android6_all_activities_lt}, \ref{fig:samsungs6edge_all_activities_lt}, \ref{fig:huaweip8_all_activities_lt}). Moreover, we also observed increases in memory consumption of key processes of the Android OS (\cfr \figurename~\ref{fig:android6_pss}, \ref{fig:huawei_pss}, \ref{fig:samsung_pss}).

\vspace{5pt}
\noindent
\textbf{(RQ1.1) Is software aging widespread across Android devices from different Android vendors?} 
In our analysis, we considered devices from four leading Android vendors (Samsung, Huawei, LG, and HTC) to address this question. We found that software aging occurs consistently across these four vendors, thus highlighting that software aging is not limited to specific Android devices (\cfr \tablename~\ref{ANOVAassumptions}, \figurename~\ref{fig:android6_all_activities_lt}). We have found that software aging effects are exacerbated by the specific Android vendors, as they apply customizations to the basic Android OS. Moreover, the workload is another factor that significantly contributes to the extent of software aging (\eg by stressing the parts of the Android OS that are affected by aging-related bugs), as in our tests the Chinese applications revealed higher software aging trends of the Android OS.

\vspace{5pt}
\noindent
\textbf{(RQ1.2) Is software aging widespread across different versions of the Android OS?}
Similarly, we analyzed the presence and the variability of software aging across different versions of the Android OS, by considering the three most recent Android releases at the time of writing. We found that all the considered Android versions are affected by software aging, thus pointing out that this problem is not limited to specific versions, but that the problem permeates the Android OS. Moreover, our tests did not show an improvement of the Android OS over time, as the most recent Android release shows aging trends that are comparable to the previous one (i.e., there are no statistically-significant differences, \cfr \figurename~\ref{fig:samsungs6edge_all_activities_lt}, \ref{fig:huaweip8_all_activities_lt}). This finding remarks the need for more extensive tests to fix aging-related bugs, and for software rejuvenation solutions to mitigate the effects of the (unavoidable) aging-related bugs that get shipped with the products.

\vspace{5pt}
\noindent
\textbf{(RQ2) Where are the software aging problems that affect the Android OS located?}
We point out that the software aging effects can be traced back to the most basic elements of the Android OS. The extent of software aging varies with the version of the Android OS, the workload and other conditions; however, the fact that software aging manifested consistently across the experiments suggests that the aging issues are not limited to components exercised by specific use cases, but are part of the fundamental components of the OS that are always exercised by the user.
We analyzed a set of metrics inside the Android OS, in order to gain insights on where and how to mitigate software aging. We found that the software aging trends are accompanied by a statistically-significant increase in memory consumption of key Android processes (\cfr \figurename~\ref{fig:android6_pss}, \ref{fig:huawei_pss}, \ref{fig:samsung_pss}). In particular, we found that the memory consumption of the \emph{System Server} (an Android OS process that runs many of the basic services of this system, including the \emph{Activity Manager} for starting user applications) is significantly correlated with the performance degradation trends (\cfr \tablename~\ref{table:perprocessmemory}). 

We then analyzed more in detail these processes, by looking at garbage collection and task-level metrics. The analysis pointed out that the \emph{System Server} spends more and more time on garbage collection during the experiments, which is an indicator that the memory utilization becomes more fragmented and ``bloated'' (i.e., burdened by unnecessary objects) over time (\cfr \figurename~\ref{fig:gc_rank}). Moreover, the task-level analysis identified the subsystems of the Android framework that exhibit increasing trends in terms of CPU utilization and virtual memory, which points out that these components are the ones most exercised in our experiments (\cfr \figurename~\ref{fig:tasks_rank}) and represent candidates for further investigation by developers (\eg to mitigate aging effects through software rejuvenation).

We applied the insights from this analysis to perform an additional experiment, in order to assess the hypothesis that aging was influenced by memory bloat in Java containers, and to get more insights about the root causes of the aging. We applied a rejuvenation mechanism in the System Server, by flushing the contents of selected Java containers. We inspected the source code of the services identified by our previous analysis, in order to locate Java containers that could benefit from software rejuvenation. The experimental results confirmed that the containers were indeed affected by aging, and that the analysis was useful to identify good targets for software rejuvenation. In principle, the same analysis could be performed on new releases of Android to trace back software aging issues, and to introduce software rejuvenation where necessary.

\section{Threats to validity}

\textbf{Selection of Android versions and device vendors.} 
As for any empirical study, the experiments can only cover a limited number of devices. We focused on three major versions of the Android OS and four popular Android devices, as discussed in Section~\ref{subsec:factors}. 
Since the Android OS running on commercial devices is not the same of the open-source Android, but it is modified to include proprietary customizations, we could not perform a diversity analysis as proposed by \citep{nagappan2013diversity}, which would require detailed information about the number of developers, the number of changes, the amount of source code, etc., which is not publicly available. 
Therefore, in order to achieve a reasonable diversity of the analyzed devices, we relied on devices from different competing vendors on the market.

\textbf{Number of experimental runs.} 
We clarify that our results are based on the analysis of a single time series, \ie a single experiment for each considered combination. 
To have more reliable values, the trends have been checked by a robust trend detection procedure. Beside the MK test, we applied three further statistical tests for trend detection, claiming a trend only if the MK test and two further tests rejected the null hypothesis of no trend. In the average, in 69 out of 72 experiments at least three out of four tests provided the same outcome, confirming the reliability of the trend detection procedure against false positives/negatives. 
The experiments had a 6-hours duration, with samples collected every 30 seconds. We used the Sen's procedure on these data to estimate the probability distribution of the trend value, to mitigate the threat of not having repetitions; this estimate represents a theoretical derivation valid under the assumptions of the Sen's procedure. Although the 95\%-confidence intervals for the estimated trends are small, replications would provide more actual values for the trend, and would empirically confirm/reject the theoretical trend's probability distribution. Due to the high cost of repetitions, we opted to spend experimental time to cover more configurations (our design has 72 experiments) and to have stable trends (by 6-hours duration). Even a few repetitions would require a significant amount of additional resources, not only in terms of time but also in terms of hardware, without significantly impacting the result.  

\textbf{Duration of the experiments.} 
The limited duration of the stress tests is an inevitable limitation of any experimental work. We calibrated the duration of our experiments to 6 hours according to data from our previous work \citep{cotroneo2016software} since, in retrospect, this duration would have been sufficient to detect all of the aging trends. There is still some risk that this duration has not been enough to point out aging trends for some of the new conditions tested in this work. However, this duration still gives us confidence that we covered the most severe trends in terms of high performance degradation rate and short time-to-aging-failure, since any missed trend would have a much longer time-of-aging-failure and smaller performance degradation.

\textbf{Workload definition and generation.} 
The workload makes use of a number of applications widely used in both the Chinese and European markets.  These include information and content applications (e.g., videos, music, camera, gallery), communication applications (e.g., browsers, email, IM), and basic applications (e.g., contacts, calendar, phone). 
Not including some types of apps, such as gaming ones, limits the generality of the conclusions: the reader should be aware of this aspect when generalizing the reported numerical trends if their workload is different from ours.

Moreover, the workload is characterized by fixed configurations of the \textit{EVENTS} factor, where one out of the three types of event is predominant (e.g., motions). Indeed, the types of events can be mixed in several other ways, such as making a workload dominated by two types of events, possibly obtaining different results. On the other hand, we can also conjecture, based on the obtained results, that changing the type of events is expected to have a little impact: in fact, the factor \textit{EVENTS} turned out to be not statistically significant in our analysis, hence we expect  that changing the percentages would have little impact.

To generate the workload, we adopted the \emph{Monkey} tool, which generates pseudo-random streams of user events. Hence the final result of a single run could be affected by the \textit{seed} selected for the pseudo-random event generation, as the seed determines the sequence of events. On the other hand, it should be noted that the possible occurrence of aging symptoms in a given configuration (that is, an experiment) does not depend on a single UI event, but on the cumulative effect of highly numerous events: the duration (6 hours) and the number of events (an event each 500ms) are so large that, given a fixed percentage of events of each type, the impact of the ordering of their execution becomes negligible. 
Indeed, a sensitivity analysis aimed at investigating the impact of the time between two consecutive events (500ms) and/or of the experiment duration on the extent of aging manifestation is an interesting future direction to study the phenomenon accounting for these factors too.

About the events generation, it should be noted that the effect of the randomly generated events could also interact with the effect caused by the other studied factors. The ANOVA we conducted on the controlled factors is a ``one-way'' ANOVA: this means that the possible effects of the interactions among multiple factors are not accounted for. This is a typical approach, since interactions usually explain a small part of variation of results. Conducting further experiments to run a two-way or \emph{n}-way ANOVA could point out such interactions.

We also warn the reader against considering the reported aging trends as the actual values expected in operation. The adopted workload (\ie the chosen apps, the time between events, etc.) is not meant to be representative of some user behaviour, because, in order to observe and clearly distinguish the possible aging effects (from the workload), a fixed and stressful workload is preferred. This has the goal of highlighting possible aging problems, so as to perform a root cause analysis and debugging. Hence, results highlight that there are indeed potential aging problems, but the real trends will depend on real workload, hence on the user behaviour.


\section{Lessons learned and conclusion}

According to the results of our experimental analysis, we derive the following lessons learned and recommendations on mitigating aging for researchers and developers.

Performing stress tests, as in our experiments, is useful for developers as a starting point to \textbf{localize and fix software aging issues in the source code of the Android OS}. For example, the processes and services with anomalous GC times and CPU/virtual memory utilization should be scrutinized with more emphasis, by performing a detailed memory profiling of these services to identify stale objects and other bad memory management patterns. Since profiling is a heavy activity and since the Android OS has an extensive code base (several millions of lines of code), it is important to restrict profiling to the components that are more prone to aging-related issues.

Another approach is to apply software rejuvenation strategies, since the underlying software aging bugs may be too costly to be fixed (e.g., due to the long time that is needed to reproduce them). Our stress tests show that \textbf{experimental data can point out aging-prone components} that are causing performance degradation and resource utilization, and that are candidates for rejuvenation actions. Since software rejuvenation comes with a cost (e.g., in terms of periods of time during which the device is not usable, or slowed-down, due to rejuvenation), it is important to keep low this cost in order to improve the quality of experience of Android users, by targeting rejuvenation on selected components. 
Knowing the Android processes most prone to software aging (such as the System Server) allows developers to focus engineering efforts to implement rejuvenation actions at the process-level, such as to introduce mechanisms to preserve any critical state during the restart of a process, such that no Android applications or other system processes are affected.

In our analysis, we presented an experiment with a simple micro-rejuvenation mechanism that periodically flushes selected Java containers to counteract memory bloat, without restarting neither the Android device nor the process that runs the service. The experimental results showed that the Android OS can indeed benefit from rejuvenation in terms of less degradation of performance. Moreover, the analysis of software aging symptoms (e.g., CPU and memory utilization by process and by task, garbage collection) was proven to be useful to identify potential candidates for software rejuvenation. Such rejuvenation mechanisms are an interesting direction for future research, as we only limited to a very small subset of containers that could be safely rejuvenated with low programming efforts.

Rejuvenation can focus on the services that are most used and/or that exhibit a higher utilization CPU and memory, such as the services of the System Server and Surface Flinger that were pointed out by our analysis. The \textbf{rejuvenation actions should be triggered with a measurement-based approach}, by monitoring the memory utilization and garbage collection as we did in our tests (e.g., by sampling the PSS and collecting logs from the ART) in order detect the onset of software aging effects.

\bibliographystyle{spbasic} 
\bibliography{bibliography}

\end{document}